\documentclass[twocolumn,final,aps,prd,longbibliography,nofootinbib,amsfonts,amssymb,amsmath,superscriptaddress]{revtex4-2}

\usepackage[utf8]{inputenc}
\usepackage[T1]{fontenc}
\usepackage{xcolor}
\usepackage{tensor}
\usepackage{enumerate}
\usepackage{physics}
\allowdisplaybreaks[1]
\usepackage{url}
\usepackage{hyperref}\hypersetup{%
	pdfauthor={Giulia Maniccia and Giovanni Montani and Marco Tosoni},%
	pdftitle={Analyzing the influence of graviton fluctuations on the inflationary spectrum with a Kucha\v{r}-Torre clock}%
}

\newcommand{\vkl}{{v_\textbf{k}^\lambda}}
\newcommand{\phik}{{\phi_\textbf{k}}}
\newcommand{\V}{{\mathcal{V}_{\lambda,\bar{\lambda}}}}

\newcommand{\Hubble}[0]{\mathrm{H}}

\renewcommand{\textcolor}[2]{#2}
\begin{document}

\title{Analyzing the influence of graviton fluctuations on the inflationary spectrum with a Kucha\v{r}-Torre clock}

\author{Giulia Maniccia}
\email{giulia.maniccia@uniroma1.it}
\affiliation{Physics Department, ``La Sapienza” University of Rome, P.le A. Moro 5, 00185 Roma, Italy}
\affiliation{Physics Department, Institute for Quantum Gravity, Theoretical Physics III, FAU Erlangen-N\"urnberg, Staudtstr. 7, 91058 Erlangen, Germany}

\author{Giovanni Montani}
\email{giovanni.montani@enea.it}
\affiliation{ENEA, FNS Department, C.R. Frascati, Via E. Fermi 45, 00044 Frascati (Roma), Italy} 
\affiliation{Physics Department, “La Sapienza” University of Rome, P.le A. Moro 5, 00185 Roma, Italy}

\author{Marco Tosoni}
\email{tosoni.1723403@studenti.uniroma1.it}
\affiliation{Physics Department, “La Sapienza” University of Rome, P.le A. Moro 5, 00185 Roma, Italy}

\date{\today}

\begin{abstract}
This paper focuses on the search for a coherent and consistent formulation to describe quantum gravity corrections to quantum field theory.
We implement two fundamental ingredients discussed in previous analyses: on one hand, the construction of the Wentzel-Kramer-Brillouin and Born-Oppenheimer picture for the Wheeler-DeWitt theory of gravity and matter by using a reference fluid as physical clock; on the other hand, we explicitly separate in the metric field its own purely classical contribution from the quantum graviton degrees of freedom. 
This allows to derive, by the expansion in a Planckian parameter, a unitary theory properly evaluating the quantum field theory modifications in the considered regime.
More specifically, we recover at zero order the standard theory on curved space-time after averaging over the graviton sector.
The physical corrections are induced at first order by (quantum) gravitons, thought as the slow-varying component of the system, as opposed to the fast quantum matter and reference fluid. The genuine contribution due to gravitons, nonfactorizable into an independent phase, provides a coherent quantum gravity modification on quantum field theory.
We show this by determining the predicted inflationary spectrum during an exact de Sitter phase of the primordial Universe, finding a breaking of the scale invariant morphology even when the inflation potential is modeled by a cosmological constant term. 
Remarkably, the behavior of such non-scale-invariant spectrum overlaps predictions previously obtained in literature by neglecting nonunitary terms emerging in those formulations.
In this respect, the present model provides a satisfactory regularization of such approaches by virtue of a more realistic construction of the physical clock for the total (gravity + matter) quantum dynamics.
\end{abstract}

\maketitle

\section{Introduction}

One of the most intriguing points addressed by proposals to quantize the gravitational field \cite{bib:thiemann-book,bib:dewitt1-1967,*bib:dewitt2-1967,*bib:dewitt3-1967} is the so-called frozen formalism \cite{bib:Isham}. This feature, very distinctive of the gravitational interactions with respect to other quantum fields, is a direct consequence of the diffeomorphism invariance \cite{bib:kuchar-1981}. 
Thus, although related to the geometrical nature of the gravitational field, this property is not directly associated to the conceptual problem of quantizing the space-time structure and it opens a related subtle point: 
how can we recover the quantum field theory on a classical curved space-time from the global quantum gravity picture in the presence of that field?
The appropriate limit can be constructed by developing the gravitational field dynamics in a Wentzel-Kramer-Brillouin (WKB)-like scenario, which allows to recover the matter field only as the intrinsically quantum structure. 

In this respect, one of the first approaches aiming to reconstruct the quantum field theory on a curved background from the Wheeler-DeWitt (WDW) equation was presented in \cite{bib:rubakov-lapchinsky-1979}. There a Tomonaga-like notion of time was introduced and a functional Schr\"odinger-like scenario properly emerged for the matter field only. 
A more convincing formulation of the same problem has been provided by \cite{bib:Vilenkin} in a rather general minisuperspace context. The separation of the system into a quasiclassical (WKB expanded) background on which a ``small'' quantum system lives, is a general framework in which the quantum dynamics of matter field can be easily accommodated. 
The crucial point here is the way to reintroduce a time coordinate in the dynamics of the matter: a label time presence comes out from the dependence of the matter wave functional on the classical gravitational variables (in turn functions of the label time itself). 
This scenario has been generalized in \cite{bib:kiefer-1991}, where the expansion parameter has been identified in a Planck-like scale, instead of $\hbar$. 
For a discussion of the equivalence of the two approaches above, see \cite{bib:digioia-2021}, where it was also emphasized how the nonunitarity emerging at first order in the quantum gravity corrections cannot be overlooked, calling attention for a different formulation of the physical clock.
By other words, while the ideas of \cite{bib:Vilenkin,bib:kiefer-1991} for a time reconstruction provides a correct answer to deal with quantum field theory on a classical geometry from the WDW equation, it certainly fails in determining the quantum gravity corrections to such a quantum field theory (QFT) reconstructed formulation \cite{bib:review-mary-2022}. 

An alternative proposal for the time definition versus the QFT limit has been outlined in \cite{bib:digioia-2021}, where the so-called ``kinematical action'' \cite{bib:kuchar-1981} has been added to the general formulation of the gravitational field. 
A systematic reformulation of the problem, on the base of the so-called ``Kucha\v{r}-Torre'' \cite{bib:KucharTorre} reference frame fixing, stands in \cite{bib:maniccia-montani-2022}. 
In this work, the similarities with the kinematical action formulation have been emphasized and a cosmological implementation of the model has been provided. This recasting of the WDW equation via a sort of Born-Oppenheimer separation into a 
``slow'' gravitational component plus a ``fast'' contribution (i.e. the matter field and the reference fluid) reinstated a unitary formulation for the matter dynamics in the presence of quantum gravity corrections. 
However, the implementation of this theory to the determination of the inflationary spectrum \cite{bib:maniccia-montani-torcellini} demonstrated that the obtained generalized QFT theory always contains a solution for which quantum gravity effects are removed, simply because they emerge as a phase factor. 

This result suggested the necessity to reconsider the way in which the real degrees of freedom of the theory must be weighted in the proposed model. 
In fact, in \cite{bib:maniccia-montani-antonini}, it has been clarified the necessity to assign to the WKB quantum part of the gravitational field a specific identity, i.e. separating the background metric into a purely classical background (the Bianchi I cosmology) plus quantum (although slow) graviton degrees of freedom. 
The important statement of this study has been the necessity to somehow average over the graviton sector before we can really speak of reconstructing a QFT model. 

In the present paper, we combine together the idea of defining a physical clock via the Kucha\v{r}-Torre fluid with the classical-quantum separation of the gravitational background.
We first construct the theory up to zeroth order in the Planckian parameter, in which QFT can be identified as the result of an averaging procedure over the graviton sector. As we will motivate, this level presents an exact coincidence between the (WKB expanded) graviton Wheeler-DeWitt equation and the needed Born-Oppenheimer gauge condition \cite{bib:kiefer-1991,bib:bertoni-finelli-venturi-1996}. 
Then we analyze the next order of approximation, at which the quantum gravity corrections on QFT manifest themselves. Also adopting in this step an appropriate gauge to eliminate some contributions from the quantum dynamics, we arrive to a new Schr\"odinger-like equation which has the simultaneous merit of being fully unitary and not presenting the graviton contributions as a simple phase factor anymore.
In this step the graviton dependence of QFT cannot be removed by a natural average procedure, i.e. it is no longer possible to get a readable equation in the average matter wave function. 
This fact leads us to postulate the necessity of a phenomenological average procedure, i.e. performed on the solution of the dynamical system instead of on its formulation. 

As the fundamental result of this study, we implement the constructed theory to the calculus of the inflationary spectrum during a pure de Sitter phase \cite{bib:weinberg-grav-cosm,bib:PrimCosm}. 
We arrive to determine quantum gravity corrections to the scale-invariant spectrum, which have the same morphology already obtained in \cite{bib:brizuela-kiefer-2016-desitter,bib:venturi-2013-spettro,bib:venturi-2015-spettro-confronto} but now we can deal with a unitary physical prescription for these corrections. 

Thus, the present analysis is a first concrete and predictive step versus the setup of a revised QFT scenario, amended for the presence of small ripples in the gravitational field, due to a relic of their quantum nature. 
Despite the difficulty in experimentally detecting such very small effects, the possibility to deal with a predictive prescription for their evolution is to be regarded as a basic step in our knowledge of quantum gravity.

The structure of the paper is as follows. In Sec. \ref{sec2:KT}, we review Kucha\v{r} and Torre's parametrized procedure for the Gaussian reference fluid clock. In Sec. \ref{sec3:gravitons-vilenkin} we revisit the treatment of graviton perturbations using the WKB approach in a Bianchi I universe. This is subsequently integrated with the Kucha\v{r}-Torre clock proposal in Sec. \ref{sec4:fluid-gravitons}. We then employ this combined framework in Sec. \ref{sec5:fluid-gravitons-spectrum} to calculate the power spectrum corrections arising from quantum gravity effects for the inflationary universe. Discussion and concluding remarks are presented in Sec. \ref{sec:conclusions}.

\section{The Kucha\v{r}-Torre clock}\label{sec2:KT}
In this section we shall review the procedure exposed in \cite{bib:KucharTorre} and \cite{bib:maniccia-montani-2022}, by means of which the reference fluid concept is exploited to recover a time definition and determine quantum gravity corrections to the matter dynamics.

The first step lies in identifying, among all the metric components, those ones that represent the physical degrees of freedom; namely those which describe the dynamical evolution of the gravitational field. This is obtained by means of a frame-fixing procedure, i.e. choosing as coordinates the Gaussian ones $(T,X^i)$ for which the metric $\gamma_{\alpha \beta}$ satisfies the following conditions:
\begin{equation}
\gamma^{00}=1, \qquad \gamma^{0i}=0\,,
\label{gammacond}
\end{equation} 
with signature $(+,-,-,-)$ and latin letters labeling spatial indices. 

In absence of matter (see \cite{bib:KucharTorre}), the conditions \eqref{gammacond} can be imposed by 
inserting an additional term in the total action of the system:
\begin{equation}
S = S^g + S^f,
\end{equation}
where $S^g$ is the Einstein-Hilbert action and 
\begin{equation}
S^f = \int d^4x \left[ \frac{\sqrt{-\gamma}}{2} \left( \gamma^{00} - 1 \right)\mathcal{F}  + \sqrt{-\gamma}\gamma^{0i} \mathcal{F}_i  \right].
\end{equation}
Here $\mathcal{F}$ and $\mathcal{F}_i$ are Lagrange multipliers, so that requiring $S$ to be stationary under their variation reproduces exactly \eqref{gammacond}. 

This additional term acts as a source in Einstein's equations, breaking the diffeomorphism invariance. Such property can however be restored via a reparametrization of the Gaussian coordinates as functions of some arbitrary coordinates $x^{\alpha}$ with metric $g_{\alpha \beta}$, namely
\begin{equation}
\begin{split}
S^f = \int d^4x &\left[ \frac{\sqrt{-g}}{2} \mathcal{F}\; \left( g^{\alpha\beta}\partial_\alpha T(x) \partial_\beta T(x) - 1 \right)\right. \\
&\left.+ \sqrt{-g} \mathcal{F}_i \;g^{\alpha\beta} \partial_\alpha T(x) \partial_\beta X^i(x) \vphantom{\frac{1}{2}} \right].
\end{split}
\end{equation}
It can be shown \cite{bib:KucharTorre}, by defining
\begin{equation}
U^\alpha = g^{\alpha\beta} \partial_\beta T, \qquad \mathcal{F}_\alpha = \mathcal{F}_i \partial_\alpha X^i,
\end{equation}
that the energy-momentum tensor associated to $S^f$ takes the following form:
\begin{equation}
T^{\alpha\beta} = \mathcal{F}U^\alpha U^\beta + \frac{1}{2}\left( \mathcal{F}^\alpha U^\beta + \mathcal{F}^\beta U^\alpha \right)
\end{equation}
i.e., a heat-conducting fluid having four velocity $U^\alpha$, energy density $\mathcal{F}$ and heat flow $\mathcal{F}_\alpha$, which we refer to as reference fluid.

By means of the 3+1 Arnowitt-Deser-Misner (ADM) spacetime foliation \cite{bib:ADM} the Hamiltonian description of the reference fluid is found:
\begin{gather}
\label{2.Hf}
H^f = W^{-1}P + W W^k P_k, \\
H_i^f = P \partial_i T + P_k \partial_i X^k,
\end{gather}
where
\begin{equation}
W = \left( 1 - h^{ij} \partial_i T \partial_j T \right)^{-1/2},\quad
W^k = h^{ij} \partial_i T \partial_j X^k,
\end{equation}
with $h_{ij}$ being the induced metric on the spatial hypersurfaces $\Sigma$ of the foliation, and $P$ and $P_k$ are the conjugate momenta of $T$ and $X^k$. 

In the presence of matter, namely a test scalar field $\phi$, the super-Hamiltonian and supermomentum constraints for the total system read as follows:
\begin{gather}
\Bigl( \hat{H}^g + \hat{H}^m + \hat{H}^f \Bigr) \Psi = 0, \label{Htot}\\
\Bigl( \hat{H}_i^g + \hat{H}_i^m + \hat{H}_i^f \Bigr) \Psi = 0, \label{Hitot}
\end{gather}
where the Dirac prescription for constraint quantization has been implemented. However, in its general formulation, the reference fluid approach does not admit a classical limit since its energy conditions are not satisfied. In fact, such conditions in the heat-conducting case require a precise relation between the (independent) multipliers $\mathcal{F}$ and $\mathcal{F}_i$, which can also be violated during the dynamical evolution \cite{bib:KucharTorre}. Such problem can be overcome by regarding the reference fluid as a quantum component, i.e. emerging at the next order in a a WKB expansion of the system in the parameter $1/M$, where
\begin{equation}
M = \frac{c^2}{32\pi G} = \frac{c m_{Pl}^2}{4\hbar}.
\label{2.M}
\end{equation}
In this way, the zeroth order represents the pure gravity limit and there is no classical limit for the reference fluid \cite{bib:maniccia-montani-2022}.

In what follows, we will also use a Born-Oppenheimer separation of the dynamics between the gravitational variables and the quantum matter ones:
\begin{equation}\label{2.psi-separated}
\Psi(h_{ij},\phi,X^\mu) = \psi(h_{ij})\chi(\phi,X^\mu;h_{ij})
\end{equation}
where $X^\mu = (T,X^i)$ and the matter wave functional $\chi$ parametrically depends on the background metric. In this scenario, the gravitational degrees of freedom are treated as slow variables, while the quantum matter ones and the reference fluid are regarded as fast.
The WKB expansion up to $\order{M^{-1}}$ of the wave functional brings
\begin{equation}
\Psi(h_{ij},\phi,X^\mu)=e^{\frac{i}{\hbar}(MS_0+S_1+\frac{1}{M}S_2)} e^{\frac{i}{\hbar} (Q_1 + \frac{1}{M} Q_2)}\,,
\label{WKB}
\end{equation}
where the $Q_n(\phi,X^\mu;h_{ij})$, $S_n(h_{ij})$ are complex functions representing, respectively, the matter and gravity sectors ($n$ labels the order of expansion). We also impose an adiabatic condition
\begin{equation}
\frac{\delta Q_n}{\delta h_{ij}} = \order{\frac{1}{M}}\,,
\label{2.adiab}
\end{equation}
and require the matter sector to live at a smaller energy scale (in terms of expectation values on the associated states) with respect to the gravitational one, namely
\begin{equation}
\frac{\langle \hat{H}^m \rangle_\chi}{\langle \hat{H}^g \rangle_\psi} = \order{\frac{1}{M}}\,.
\label{2.energy}
\end{equation}
The latter illustrates why the matter component lacks terms $\propto M$ in \eqref{WKB} -- the leading order corresponds to pure gravity. Therefore, we can also consider any matter effect at the Planck scale to be negligible, so that the gravitational wave functional must satisfy the following:
\begin{gather}
\hat{H}^g \psi(h_{ij}) = 0, \label{Hg}\\
\hat{H}_i^g \psi(h_{ij}) = 0. \label{Hig}
\end{gather}
The complete set of constraints is therefore given by Eqs.~\eqref{Htot}--\eqref{Hitot} and \eqref{Hg}--\eqref{Hig}.

The expansion in $M^{-1}$ brings a hierarchy of constraints, which we briefly recall. At first order i.e. $\order{M}$ one has
\begin{gather}
\frac{1}{2} \left( \frac{\delta S_0}{\delta h_{ij}} \right)^2 + V(h_{ij}) = 0, \\
-2h_{ij} D_k \frac{\delta S_0}{\delta h_{kj}} = 0,
\end{gather}
being $D_k$ the covariant derivative over $\Sigma$ and $V(h_{ij})$ the gravitational potential coming from the Einstein-Hilbert action. 
These expressions can be recognized as the Hamilton-Jacobi equation for gravity and the diffeomorphism invariance condition for $S_0$.

At $\order{M^0}$ one obtains, combining the super-Hamiltonian and supermomentum contributions
\begin{equation}
\begin{split}\label{2.Schrod-t0}
    \hat{\mathcal{H}}^m & \chi_0 = \int_\Sigma d^3x \left( N \hat{H}^m + N^i \hat{H}_i^m \right) \chi_0 \\
    & = i \hbar \int_\Sigma d^3x \left[ N \left( W^{-1} \frac{\delta}{\delta T} + W W^k \frac{\delta}{\delta X^k} \right)\right. \\
    & \left.\ \ + N^i \left( (\partial_i T) \frac{\delta}{\delta T} + (\partial_i X^k) \frac{\delta}{\delta X^k} \right) \right] \chi_0 \\
    & \equiv i \hbar \frac{\delta}{\delta \tau} \chi_0
\end{split}
\end{equation}
with $\chi_0 = \exp(iQ_1/\hbar)$. \textcolor{blue}{Here, the constraint equations allowed us to establish an equivalence between the matter Hamiltonian $\hat{\mathcal{H}}^m$, written in terms of the corresponding super-Hamiltonian and supermomentum, and an expression involving functional derivatives in the fluid variables. This last term has been interpreted as a time derivative operator: its definition corresponds to recasting the matter dynamics in a functional Schr\"{o}dinger form.}

The next order $M^{-1}$ can be summed with $\order{M^0}$ to obtain
\begin{equation}\label{2.schrod-correct}
    i\hbar \frac{\delta}{\delta \tau} \chi_1 = \hat{\mathcal{H}}^m \chi_1 + \hat{\mathcal{H}}^{QG}\chi_1,
\end{equation}
where 
\begin{equation}
    \hat{\mathcal{H}}^{QG} = 2i\hbar \int_\Sigma d^3x \left[- N \frac{\delta S_0}{\delta h_{ij}} \frac{\delta}{\delta h_{ij}} + N^i h_{ij} D_k \frac{\delta}{\delta h_{kj}} \right].
\end{equation}
Comparing with \eqref{2.Schrod-t0}, Eq.~\eqref{2.schrod-correct} describes a modification to the Schr\"{o}dinger evolution for the $\order{M^{-1}}$ matter wave functional $\chi_1 = \exp\left[i(Q_1+ M^{-1}Q_2)/\hbar\right]$.
A crucial property of the corrective terms in Eq.~\eqref{2.schrod-correct} stands in their unitarity, following from the classical Hamilton-Jacobi solution $S_0$ and the (Hermitian) conjugate momenta $\Pi_{ij}$ \cite{bib:maniccia-montani-2022}, thus overcoming the nonunitarity plaguing the next-level dynamics of Vilenkin's proposal \cite{bib:digioia-2021}.

\section{Graviton perturbations in the WKB approach}\label{sec3:gravitons-vilenkin}

The time parameter proposed by Vilenkin \cite{bib:Vilenkin} makes use of a separation between semiclassical and quantum degrees of freedom: 
the quantum subsystem is identified by variables $q_a$ ($a=1,\ldots m$), while the minisuperspace semiclassical degrees of freedom are $h_A$ ($A=1,\ldots N$) in analogy with the gravity-matter separation of \eqref{2.psi-separated}. This allows for a WKB expansion in $\hbar$ of the semiclassical sector:
\begin{equation}
\Psi(h_A,q_a) = A(h_A)e^{\frac{i}{\hbar}S(h_A)}\chi(q_a;h_A),\label{3.psiVil}
\end{equation}
where $A(h_A)$ and $S(h_A)$ are real-valued functions; this semiclassical component can be identified with gravity \cite{bib:digioia-2021}. 

In a diagonal minisuperspace model, we can take $N^i=0$ so that the supermomentum constraint is identically satisfied \cite{bib:brizuela-kiefer-2016-desitter}. We then deal with the super-Hamiltonian constraints only, i.e., the WDW equations of the full system and the gravitational sector alone:
\begin{gather}
\left( \hat{H}^g + \hat{H}^m \right) \Psi(h_A,q_a) = 0, \label{3.VilWDW}\\
\hat{H}^g A(h_A)e^{\frac{i}{\hbar}S(h_A)} = 0.\label{3.gravWDW}
\end{gather}

Expansion in powers of $\hbar$ allows to reproduce at $\order{\hbar^0}$ the Hamilton-Jacobi equation for the gravitational field:
\begin{equation}
    G_{AB}\frac{\partial S}{\partial h_A} \frac{\partial S}{\partial h_B} + V(h_A) = 0,
    \label{VilHJ}
\end{equation}
where $G_{AB}$ and $V$ are the minisupermetric and the gravitational potential. At $\order{\hbar}$, Eqs. \eqref{3.VilWDW}--\eqref{3.gravWDW} bring
\begin{gather}
G_{AB} \frac{\partial}{\partial h_A} \left( A^2 \frac{\partial S}{\partial h_B} \right) = 0,\\
i\hbar \partial_\tau \chi = N \hat{H}^m \chi. \label{VilSchr}
\end{gather}
We can easily recognize \eqref{VilSchr} as a functional Schr\"odinger equation for the matter dynamics, in which
\begin{equation}
\partial_\tau \chi = 2NG_{AB} \frac{\partial S}{\partial h_A} \frac{\partial}{\partial h_B} \chi = \dot{h}_A \frac{\partial}{\partial h_A} \chi
\label{Viltime}
\end{equation}
has been defined (here and the dot stands for derivation with respect to the label time). 
Remarkably, here the time dependence of the matter sector lies in the fact that $\chi$ is a functional of the $h_A$, which are themselves functions of the label time $t$. 

The ambiguities present in this approach have been highlighted in \cite{bib:maniccia-montani-antonini}. It has been argued that the presence of an amplitude $A(h_A)$ implies that the gravitational variables possess a quantum component, namely $h_A = h_A^0 + \delta h_A$ being $\delta h_A$ the quantum correction. Therefore, the definition \eqref{Viltime} must be modified as to include only the semiclassical contributions, and the dynamics of the quantum variables $\delta h_A$ must also be specified. 
Moreover, Eq.~\eqref{VilSchr} does not reproduce the limit of QFT on curved spacetime in this sense, since $\chi$ depends on the quantum corrections of the background, which should instead behave as a classical object.

The \textit{a priori} imposition of the gravitational WDW constraint \eqref{3.gravWDW} poses another point of discussion: such requirement breaks the natural gauge invariance of the Born-Oppenheimer separation associated to opposite phase shifts of the gravitational and matter components.

In order to overcome these ambiguities, the formulation in \cite{bib:maniccia-montani-antonini} proposed to incorporate quantum corrections (in the form of cosmological perturbations) for the gravitational background, which was chosen as the vacuum diagonal Bianchi I model. Such anisotropic setting still allows, in absence of matter backreaction, to decouple scalar and vector perturbations \cite{bib:hu-1978,bib:miedema-1993}; because of that, only tensor gravitational fluctuations were taken into account.

Let us briefly recall the formalism and results of \cite{bib:maniccia-montani-antonini}: we start with a system composed of the (classical) Bianchi I geometry, tensor perturbations of this background and a matter scalar field $\varphi$ that is the only source of scalar perturbations; both kinds of fluctuations are described by means of the Mukhanov-Sasaki formalism \cite{bib:Mukhanov,bib:brizuela-kiefer-2016-desitter}, i.e. with the Fourier modes of the corresponding gauge-invariant variables $v_\textbf{k}^\lambda$ and $\phi_\textbf{k}$. 

The minisuperspace background is characterized by the Misner variables $\alpha$ (the logarithmic volume of the Universe) and $\beta_\pm$ (representing the anisotropies) \cite{bib:Gravitation,bib:PrimCosm}:
\begin{equation}
    \textcolor{blue}{ds^2 = N^2(t)dt^2 - e^\alpha (e^\beta)_{ij} dx^i dx^j}
\end{equation}
where $\beta=\mathrm{diag}(\beta_++\sqrt{3}\beta_-,\beta_+-\sqrt{3}\beta_-,-2\beta_+)$ is traceless \textcolor{blue}{and the choice of signature reflects Sec.~\ref{sec2:KT}}. The corresponding super-Hamiltonian operator is
\begin{equation}
    \hat{H}^I = \frac{4}{3M} e^{\frac{3}{2}\alpha} \left[ \textcolor{blue}{\partial_+^2 + \partial_-^2  - \partial_\alpha^2} \right],
    \label{3.HI}
\end{equation}
being $M$ the Planckian parameter defined in (\ref{2.M}).

The contribution of tensor perturbations of the metric, i.e. gravitons, brings in the conformal gauge $N=e^\alpha$ (\textcolor{blue}{see \cite{bib:Mukhanov} for the original derivation in a homogeneous setting, \cite{bib:brizuela-kiefer-2016-desitter} for a de Sitter universe and \cite{bib:PPU} for the anisotropic case)}:
\begin{equation}
    \hat{H}^{v^\lambda} = \frac{e^{-\alpha}}{2} \sum_{\textbf{k},\lambda} \left[ -\hbar^2 \partial_\vkl^2 + \omega_{k,\eta}^2 (\vkl)^2 + \V \right]\,,
    \label{3.Hv}
\end{equation}
where $\eta$ is the conformal time. Therefore, the Fourier modes behave as time-dependent harmonic oscillators with frequency $\omega_{k,\eta}^2=k^2-z''_\lambda/z_\lambda$, with $z_\lambda(\eta,k_i)$ being a function of the Bianchi I metric and polarization state \cite{bib:PPU} and $'=\partial_\eta$. In the spirit of \cite{bib:maniccia-montani-antonini}, the potential term $\V$ represents the effect of anisotropies by means of which a mixing between polarization states can occur \cite{bib:PPU}. We stress that, while in principle anisotropic backgrounds admit a coupling between tensor and scalar modes \cite{bib:PPU} akin to a seesaw mechanism, here such effect is suppressed by the assumption that the matter backreaction is negligible. 

The fluctuations $\phi \equiv \delta \varphi$ of the scalar field $\varphi$ are the only possible source of scalar perturbations for the classical background, whose contribution in Fourier space reads 
\begin{equation}
    \hat{H}^\phi = \frac{e^{-\alpha}}{2} \sum_\textbf{k} \left[ -\hbar^2 \partial_\phik^2 + \nu_{k,\eta}^2 (\phik)^2 \right],
    \label{3.Hphi}
\end{equation}
where now the oscillator frequency is $\nu_{k,\eta}^2=k^2-(e^{\alpha})''/e^{\alpha}$ \cite{bib:Mukhanov,bib:brizuela-kiefer-2016-desitter,bib:PPU}.

Thus, we deal with the total WDW constraint
\begin{equation}
\hat{H} \Psi = \left( \hat{H}^I + \hat{H}^{v^\lambda} + \hat{H}^\phi \right) \Psi = 0\,.
\label{3.WDW}
\end{equation}
\textcolor{blue}{Here we are taking into account both the background and the perturbations sector within the WKB picture: the former component will enter the Hamilton-Jacobi function corresponding to the classical limit [i.e. $S_0$, see Eqs.~\eqref{5.HJ}, \eqref{5.S0independence}], while the latter one is inherently quantum. Therefore the gravitons sector will be averaged over to provide the effective matter evolution on the given background in the QFT sense. Let us assume} a Born-Oppenheimer separation for $\Psi$ in the form
\begin{equation}\label{3.psi-separation}
\Psi = \psi(\alpha,\beta_\pm,v_\textbf{k}^\lambda) \ \chi(\phi_\textbf{k};\alpha,\beta_\pm,v_\textbf{k}^\lambda)
\end{equation}
in the hypothesis of negligible backreaction, backed by the different energy scales of gravity and matter.

Equation \eqref{3.psi-separation} can now be WKB expanded in the Planckian parameter $1/M$, as in \cite{bib:kiefer-1991,bib:bertoni-finelli-venturi-1996}, which for our purpose is equivalent to the $\hbar$ expansion \cite{bib:digioia-2021} but it identifies a pure gravity limit at leading order (see Sec.\ref{sec2:KT}). 
We can write up to $\order{M^0}$:
\begin{equation}
\Psi = e^{\frac{i}{\hbar}(MS_0+S_1)}e^{\frac{i}{\hbar}Q_1}
\end{equation}
where $S_n(\alpha,\beta_\pm,v_\textbf{k}^\lambda)$ and $Q_n(\phi_\textbf{k};\alpha,\beta_\pm,v_\textbf{k}^\lambda)$ are now complex-valued functions, differently from \eqref{3.psiVil}. 
Interpreting the fluctuations $v_\textbf{k}^\lambda$ as ``slow'' quantum variables, we write $S_0(\alpha,\beta_\pm)$ as independent of them and corresponding to the classical background; \textcolor{blue}{this property can also be obtained \textit{a posteriori} from the WKB expansion, see the setup in Sec.~\ref{sec4:fluid-gravitons} and Eq.~\eqref{5.S0independence}.}

Note that here the gravitational constraint \eqref{Hg} is not imposed \textit{a priori}, so that the system is still invariant under the gauge transformation:
\begin{equation}
\psi \rightarrow e^{\frac{i}{\hbar} \theta} \psi, \qquad \chi \rightarrow e^{-\frac{i}{\hbar} \theta} \chi,
\end{equation}
where $\theta=\theta(\alpha,\beta_\pm,v_\textbf{k}^\lambda)$; moreover, we shall not implement the adiabatic condition (\ref{2.adiab}).

Writing Eq.~\eqref{3.WDW} in the conformal gauge and expanding in $1/M$, the leading order reproduces the Hamilton-Jacobi equation for the Bianchi I model 
\begin{equation}
    \textcolor{blue}{(\partial_\alpha S_0)^2 - (\partial_+ S_0)^2 - (\partial_- S_0)^2 = 0,}
    \label{5.HJ}
\end{equation}
with solution:
\begin{equation}
S_0(\alpha,\beta_\pm)=k_\alpha \alpha + k_+ \beta_+ + k_- \beta_-
\label{3.S0}
\end{equation}
where we select $k_\alpha < 0$ for an expanding Universe.

The next order $\order{M^0}$ can be written in compact form by defining $\psi_1=\exp(iS_1/\hbar)$:
\begin{equation}
\begin{split}
& \vphantom{\frac{1}{2}} -i\hbar(\partial_T \psi_1) \chi - i\hbar (\partial_T \chi)\psi_1 \\
& + \frac{1}{2} \sum_{\textbf{k},\lambda} \left[ \omega_{k,\eta}^2 (v_\textbf{k}^\lambda)^2 +\mathcal{V}_{\lambda,\bar{\lambda}} - \hbar^2 \partial_{v_\textbf{k}^\lambda}^2 \right] \psi_1 \chi \\
& + \frac{1}{2} \sum_\textbf{k} \left[ \nu_{k,\eta}^2 \phi_\textbf{k}^2 - \hbar^2 \partial_{\phi_\textbf{k}}^2 \right] \psi_1 \chi = 0
\label{3.WDWM0}
\end{split}
\end{equation}
with the following time definition
\begin{equation}
\partial_T = - \frac{8}{3} e^{- \frac{\alpha}{2}} \Bigl[ \textcolor{blue}{\partial_+ S_0 \partial_+ + \partial_- S_0 \partial_- - \partial_\alpha S_0 \partial_\alpha} \Bigr].
\end{equation}
This functional operator is analogous to \eqref{Viltime} but it is constructed from the classical degrees of freedom only.

One can now exploit the gauge freedom of the Born-Oppenheimer separation to perform a gauge fixing, thus imposing an additional condition on $\psi_1$:
\begin{equation}
\begin{split}
i \hbar & \partial_T \psi_1^* + \frac{1}{2} \sum_{\textbf{k},\lambda} \Bigl[ \omega_{k,\eta}^2 (v_\textbf{k}^\lambda)^2+ \mathcal{V}_{\lambda,\bar{\lambda}} - \hbar^2 \partial_{v_\textbf{k}^\lambda}^2 \Bigr] \psi_1^* = 0.
\end{split}
\label{3.gaugefixing}
\end{equation}
This coincides with the gravitational scalar constraint at $\order{M^0}$, which can therefore be interpreted as the result of a gauge-fixing procedure rather than an \textit{a priori} imposition \cite{bib:maniccia-montani-antonini}. Equation \eqref{3.gaugefixing} also specifies the gravitons' dynamics over the classical background.

In order to address the dependence of $\chi$ on the background perturbations in \eqref{3.psi-separation}, one can implement an averaging procedure over them.  Defining an averaged wave functional
\begin{equation}\label{3.theta-averaged}
\Theta(\phi_\textbf{k};\alpha,\beta_\pm) = \int \prod_{\textbf{k},\lambda} dv_\textbf{k}^\lambda\, 	\psi_1^*\, \psi_1 \, \chi
\end{equation}
and choosing appropriate boundary conditions for the integration, it is possible to rewrite Eq.~\eqref{3.WDWM0} (multiplying  by $\psi_1^*$ and integrating over all $v_\textbf{k}^\lambda$) as
\begin{equation}
i\hbar \partial_T \Theta = N \hat{H}^\phi \Theta,
\end{equation}
which now represents the desired limit of QFT on curved spacetime.

\section{Kucha\v{r}-Torre clock for the perturbed Bianchi I dynamics}\label{sec4:fluid-gravitons}

We now proceed to unify the models exposed in Sec.~\ref{sec2:KT} and \ref{sec3:gravitons-vilenkin}: the aim of the present work is to obtain a theory in which the reference fluid approach is applied to a background with tensor gravitational perturbations. We shall therefore deal with a system composed of a Bianchi I minisuperspace, gravitons' degrees of freedom, a test scalar matter field and the reference fluid.

Working with the diagonal Bianchi I model in the Misner variables $(\alpha,\beta_\pm)$, we choose the ADM splitting in such a way that $N^i=0$, so that the supermomentum constraint is identically satisfied (as in Sec.~\ref{sec3:gravitons-vilenkin}). 
Moreover, the spatial diffeomorphisms invariance allows us to impose the Gaussian time condition alone $\gamma^{00}=1$ in the reference fluid sector, reducing its formulation to the single fluid variable $T(x)$. 

\subsection{WDW equation of the model}
The super-Hamiltonian constraint takes contributions from each of the four sectors that compose the system; since both the classical background and the (tensor and scalar) perturbations are treated as in Sec.~\ref{sec3:gravitons-vilenkin}, their corresponding operators are given by \eqref{3.HI}--\eqref{3.Hphi}, respectively.

For the reference fluid sector, due to our choice of the ADM splitting, we shall implement the Gaussian time condition only, so that the fluid is in the form of an incoherent dust \cite{bib:KucharTorre} described by means of the coordinate $T(x)$ alone. Therefore there are no spatial momenta $P^k$ and the fluid super-Hamiltonian \eqref{2.Hf}, once the fluid variable has been properly quantized by taking $\hat{P}=-i\hbar \delta/\delta T(x)$, reads
\begin{equation}
    \hat{H}^f = - i\hbar W^{-1} \frac{\delta}{\delta T(x)}. 
\end{equation}
Since $T$ is a function of a generic coordinate system due to reparametrization, it is clear that the derivative in $\hat{H}^f$ is a functional one. The fact that this sector contains only first derivatives in the fluid variables will be of great use for the construction of a time definition from the reference fluid.

The total WDW constraint, which encodes all the dynamical information for the total wave functional $\Psi$, is thus given by:
\begin{equation}
    \hat{H}\Psi = \left( \hat{H}^I + \hat{H}^{v^\lambda} + \hat{H}^\phi + \hat{H}^f \right) \Psi = 0,
    \label{4.WDW}
\end{equation}
\textcolor{blue}{where, in analogy with Sec.~\ref{sec3:gravitons-vilenkin}, we are considering within the gravitational sector both the background and its tensor perturbations, playing different roles in the suggested scheme}. Following the line of thought of the previous sections, a Born-Oppenheimer separation of the dynamics between gravitational and matter variables is performed:
\begin{equation}
    \Psi=\psi(\alpha,\beta_\pm,\vkl)\chi(\phik,T;\alpha,\beta_\pm,\vkl).
    \label{4.BO}
\end{equation}
Note that the reference fluid is part of the quantum matter sector, so that the gravitational wave functional $\psi$ does not depend on it. Each sector can then be WKB expanded in the parameter $1/M$ defined by (\ref{2.M})
\begin{equation}
    \Psi = e^{\frac{i}{\hbar}(MS_0+S_1+\frac{1}{M}S_2)} e^{\frac{i}{\hbar}(Q_1+\frac{1}{M}Q_2)},
    \label{4.WKB}
\end{equation}
where $S_n(\alpha,\beta_\pm,\vkl)$ and $Q_n(\phik,T;\alpha,\beta_\pm,\vkl)$ are complex functions and we neglected terms of order $M^{-2}$ or higher. \textcolor{blue}{Let us note that here we are allowing for the lowest-order gravitational function $S_0$ to depend also on the graviton degrees of freedom, differently from Sec.~\ref{sec3:gravitons-vilenkin}; however this property will be ruled out during the expansion of the constraint equations, see Eq.~\eqref{5.S0independence}.}
This separation is once again justified by the assumption that matter and gravity live at different energy scales, namely we require again the condition~\eqref{2.energy}.

The Born-Oppenheimer separation \eqref{4.BO} naturally implies that the matter functions $Q_n(\phik,T;\alpha,\beta_\pm,\vkl)$ are at all orders weakly dependent on the gravitational degrees of freedom, so that an adiabatic condition analogous to \eqref{2.adiab} can be implemented. However, it must be modified since in this model the gravitational variables are divided into a set of classical ones $h_C=(\alpha,\beta_\pm)$ and another set of quantum ones $\vkl$; to reflect this difference and highlight the quantum nature of gravitons, Eq.~\eqref{2.adiab} is recast as
\begin{equation}
    \frac{\delta Q_n}{\delta h_C} = \order{\frac{1}{M^2}}, \quad \frac{\delta Q_n}{\delta \vkl} = \order{\frac{1}{M}}.
    \label{4.adiab}
\end{equation}
The interpretation of \eqref{4.adiab} is that all gravitational variables are slow, but the dependence of the $Q_n$ on the classical ones is even weaker than the dependence on the quantum degrees of freedom.

As noted in the previous section, imposing \textit{a priori} the gravitational WDW equation $\hat{H}^g \psi=0$ would break the gauge invariance of the Born-Oppenheimer separation \eqref{4.BO}. Thus, we will not make use of such additional constraint, which will later be recovered as the effect of a gauge fixing procedure.

The total WDW constraint can now be recast into a hierarchy of order-by-order equations in powers of $1/M$ by substituting the expanded wave functional \eqref{4.WKB} into Eq.~\eqref{4.WDW}.

\subsection{WKB expansion up to $\order{M^{-1}}$}
The leading order in the expansion of Eq.~\eqref{4.WDW} is given by $M^2$: the only contribution comes from $H^{v^\lambda}$ and reads
\begin{equation}\label{5.S0independence}
    \sum_{\textbf{k},\lambda} (\partial_{\vkl} S_0)^2 = 0.
\end{equation}
Since all $\vkl$ are independent degrees of freedom, this reduces to the requirement $\partial_\vkl S_0 = 0$, i.e., $S_0$ must be independent of tensor perturbations: $S_0 = S_0(\alpha,\beta_\pm)$, \textcolor{blue}{as assumed in Sec.~\ref{sec3:gravitons-vilenkin}}. Thus the $\order{M}$ wave functional depends on the classical background only, as was assumed from the beginning in the model of Sec.~\ref{sec3:gravitons-vilenkin}. This feature reinforces the idea that, while gravitons are indeed slow variables, they still present an inherently quantum nature that differentiates them from the classical background.

The $\order{M}$ corresponds again to the Hamilton-Jacobi Eq.~\eqref{5.HJ} of the gravitational background, which admits the solution \eqref{3.S0} for a Bianchi I minisuperspace geometry. It is then clear that the gravitational wave functional $\psi$ represents the cosmological background at leading order, i.e., the pure gravity limit since the quantum matter sector has no contribution of the same order.

At $\order{M^0}$ all sectors contribute to the WDW constraint, which in the conformal gauge reads
\begin{equation}
    \begin{split}
        & -\frac{8}{3} e^{-\frac{\alpha}{2}} \left[ \partial_+ S_0 \partial_+ S_1 + \partial_- S_0 \partial_- S_1 - \partial_\alpha S_0 \partial_\alpha S_1 \right] \\
        & + \frac{1}{2} \sum_{\textbf{k},\lambda} \left[ \omega_{k,\eta}^2(\vkl)^2 +\V - i\hbar \partial_\vkl^2 S_1 + (\partial_\vkl S_1)^2  \right] \\
        & + \frac{1}{2} \sum_\textbf{k} \left[ \nu_{k,\eta}^2(\phik)^2 -i\hbar \partial_\phik^2 Q_1 + (\partial_\phik Q_1)^2 \right] \\
        & + e^\alpha W^{-1} \frac{\delta Q_1}{\delta T} = 0.
        \label{5.WDWM0}
    \end{split}
\end{equation}
Labeling again the corresponding wave functional as
\begin{equation}\label{defps1chi1}
    \psi_1 = e^{\frac{i}{\hbar} S_1}, \qquad \chi_1 = e^{\frac{i}{\hbar} Q_1},
\end{equation}
Eq.~(\ref{5.WDWM0}) can be recast in the form:
\begin{equation}
    \begin{split}
       & \left( \frac{8}{3} i\hbar e^{-\frac{\alpha}{2}} \left[ \partial_+ S_0 \partial_+ + \partial_- S_0 \partial_- - \partial_\alpha S_0 \partial_\alpha \right] \psi_1 \right) \chi_1 \\
        & + \left( \frac{1}{2} \sum_{\textbf{k},\lambda} \left[ \omega_{k,\eta}^2(\vkl)^2 +\V - \hbar^2 \partial_\vkl^2 \right] \psi_1 \right) \chi_1 \\
        & + \psi_1 \left( \frac{1}{2} \sum_\textbf{k} \left[ \nu_{k,\eta}^2(\phik)^2 - \hbar^2 \partial_\phik^2 \right] \chi_1 \right) \\
        & + \psi_1 \left( - i\hbar \ e^\alpha W^{-1} \frac{\delta}{\delta T} \chi_1 \right) = 0.
    \end{split}
\end{equation}
Here the absence of terms coming from the action of $\hat{H}^g$ on $\chi$ is due to the condition \eqref{4.adiab}: those contributions that contain derivatives of the $Q_n$ with respect to the gravitational variables are moved to the next orders.

It is important to recall the remaining symmetry that persists in this model. The Born-Oppenheimer separation \eqref{4.BO} is invariant under local phase shifts of the gravity and matter wave functionals:
\begin{equation}
    \psi \rightarrow e^{\frac{i}{\hbar}\theta} \psi, \ \ \ \ \ \ \ \ \chi \rightarrow e^{-\frac{i}{\hbar}\theta} \chi,
    \label{5.GT}
\end{equation}
where the function $\theta=\theta(\alpha,\beta_\pm,\vkl)$ identifies the transformation. This characteristic freedom can be exploited to impose an additional gauge fixing condition; in particular, we choose to work in the gauge for which the following equation holds:
\begin{equation}
    \begin{split}
        & \frac{8}{3} i\hbar e^{-\frac{\alpha}{2}} \left[ \partial_+ S_0 \partial_+ + \partial_- S_0 \partial_- - \partial_\alpha S_0 \partial_\alpha \right] \psi_1 \\
        & + \frac{1}{2} \sum_{\textbf{k},\lambda} \left[ \omega_{k,\eta}^2(\vkl)^2 +\V - \hbar^2 \partial_\vkl^2 \right] \psi_1 = 0.
        \label{5.gaugefix0}
    \end{split}
\end{equation}
Equation \eqref{5.gaugefix0} can be recast as a differential equation for $\theta$, the solution  of which identifies the transformation from an arbitrary gauge to the one in which \eqref{5.gaugefix0} holds.

It can be shown that the gauge-fixing condition \eqref{5.gaugefix0} is the complex conjugate of Eq.~\eqref{3.gaugefixing} once the functional $\partial_T$ has been made explicit. Therefore, the physical meaning of this additional constraint is precisely the same, i.e. the gravitational WDW constraint $\hat{H}^g \psi = 0$, and the gravitational super-Hamiltonian constraint can again be seen as a gauge-fixing of the Born-Oppenheimer separation instead of an \textit{a priori} imposition that ends up breaking such symmetry. 

After the gauge fixing, the $\order{M^0}$ equation
\begin{equation}
    \frac{1}{2} \sum_\textbf{k} \left[ \nu_{k,\eta}^2(\phik)^2 - \hbar^2 \partial_\phik^2 \right] \chi_1 = i\hbar \ e^\alpha W^{-1} \frac{\delta}{\delta T} \chi_1
\end{equation}
presents on the left-hand side the term $N\hat{H}^\phi \chi_1$ (we simplified the wave functional $\psi_1$ that acts in both sides as purely multiplicative). Thus the integration over the spatial hypersurface $\Sigma$ brings
\begin{equation}
    \hat{\mathcal{H}}^\phi \chi_1 = \int_\Sigma d^3x \ N \hat{H}^\phi \chi_1 = i \hbar \int_\Sigma d^3x \ e^\alpha W^{-1} \frac{\delta}{\delta T} \chi_1.
\end{equation}
We can now define the time parameter in terms of the reference fluid
\begin{equation}
    \frac{\delta}{\delta \tau} = \int_\Sigma d^3x \ e^\alpha W^{-1} \frac{\delta}{\delta T(x)}\,,
    \label{5.time}
\end{equation}
which describes the evolution of the matter wave functional via the Schr\"odinger (functional) equation
\begin{equation}
    \hat{\mathcal{H}}^\phi \chi_1 = i \hbar \frac{\delta}{\delta \tau} \chi_1.
    \label{5.Schr}
\end{equation}
We stress that \eqref{5.time} is consistent with the definition in \eqref{2.Schrod-t0} after taking $N^i=0$, $N=e^\alpha$ and using the fluid variable $T(x)$ only.

Finally $\order{M^{-1}}$ brings
\begin{align}
    \begin{split}
        \frac{4}{3}& e^{-\frac{\alpha}{2}} \left[ i\hbar \left( \partial_+^2 + \partial_-^2 - \partial_\alpha^2 \right) S_1 - (\partial_+ S_1)^2 - (\partial_- S_1)^2 \right.\\
        &\left. + (\partial_\alpha S_1)^2 - 2 \left( \partial_+ S_0 \partial_+ + \partial_- S_0 \partial_- - \partial_\alpha S_0 \partial_\alpha \right) S_2 \right] \\
        & + \frac{1}{2} \sum_{\textbf{k},\lambda} \left[ - i\hbar \partial_{v_\textbf{k}^\lambda}^2 + 2 \partial_{v_\textbf{k}^\lambda} S_1 \partial_{v_\textbf{k}^\lambda} \right] \left( S_2 + MQ_1 \right) \\
        & + \frac{1}{2} \sum_\textbf{k} \left[ - i\hbar \partial_{\phi_\textbf{k}}^2 Q_2 + 2 \partial_{\phi_\textbf{k}} Q_1 \partial_{\phi_\textbf{k}} Q_2 \right] \\
        & + e^\alpha W^{-1} \frac{\delta Q_2}{\delta T} = 0,
        \label{5.WDWM-1}
    \end{split}
\end{align}
which is devoid of terms $\partial_{v_\textbf{k}^\lambda}Q_2$ and $\partial_{\alpha} Q_1$ due to the adiabatic conditions \eqref{4.adiab}. Now labeling 
\begin{equation}\label{5.def-psi2-chi2}
    \psi_2 = e^{\frac{i}{\hbar}(S_1+M^{-1}S_2)}, \qquad\chi_2=e^{\frac{i}{\hbar}(Q_1+M^{-1}Q_2)},
\end{equation}
Eqs.~\eqref{5.WDWM-1} and \eqref{5.WDWM0} can be recast as a single one
\begin{widetext}
\begin{equation}\label{5.WDWM-1lunga}
    \begin{split}
        & \left( \frac{4}{3} e^{-\frac{\alpha}{2}}\left[ 2i\hbar \left( \partial_+ S_0 \partial_+ + \partial_- S_0 \partial_- - \partial_\alpha S_0 \partial_\alpha \right)+ \frac{\hbar^2}{M} \left( \partial_+^2 + \partial_-^2 - \partial_\alpha^2 \right) \right] \psi_2 + \frac{1}{2} \sum_{\textbf{k},\lambda} \left[ \omega_{k,\eta}^2 (v_\textbf{k}^\lambda)^2 + \V - \hbar^2 \partial_{v_\textbf{k}^\lambda}^2 \right] \psi_2 \right) \chi_2 \\
        & + \psi_2 \left( \frac{1}{2} \sum_\textbf{k} \left[ \nu_{k,\eta}^2 \phi_\textbf{k}^2 - \hbar^2 \partial_{\phi_\textbf{k}}^2 \right] \chi_2 \right) + \psi_2 \left( -i\hbar \ e^\alpha W^{-1} \frac{\delta}{\delta T} \chi_2 \right) + \psi_2 \left( \frac{1}{2} \sum_{\textbf{k},\lambda} \left[ - \hbar^2 \partial_{v_\textbf{k}^\lambda}^2 - 2i\hbar \partial_{v_\textbf{k}^\lambda} S_1 \partial_{v_\textbf{k}^\lambda} \right] \chi_2 \right) = 0.
    \end{split}
\end{equation}
\end{widetext}
We will now show that the terms containing derivatives of $\chi_2$ with respect to the $\vkl$ [absent in the previous order due to \eqref{4.adiab}] correspond to quantum gravity corrections to the matter dynamics.

First, we recall that the time definition \eqref{5.time} has been obtained via the the gauge-fixing condition \eqref{5.gaugefix0}, i.e., with a specific choice of $\theta$. Thus it would seem that there is no residual gauge freedom for this model. However, one can expand $\theta$ in the same parameter $1/M$:
\begin{equation}
    \theta(\alpha,\beta_\pm,\vkl)=\theta_1(\alpha,\beta_\pm,\vkl)+M^{-1}\theta_2(\alpha,\beta_\pm,\vkl)
    \label{5.thetaexp}
\end{equation}
where we omitted terms of $\order{M^{-2}}$ or higher. The corresponding transformation acts as
\begin{equation}
    \psi \rightarrow e^{\frac{i}{\hbar}\theta} \psi = e^{\frac{i}{\hbar}\theta_1} e^{\frac{i}{\hbar M} \theta_2} \psi\,,
\end{equation}
and similarly for $\chi$, thus the total gauge transformation can be seen as the composition of two separate transformations, one of $\order{M^0}$ and one of $\order{M^{-1}}$. In other words, there is a gauge freedom at every order of expansion. Now observing that Eq.~\eqref{5.gaugefix0} fixed only the $\order{M^0}$ function $\theta_1$, we can require an additional condition at the next order for $\theta_2$, namely the one corresponding to the gravitational WDW constraint up to $\order{M^{-1}}$:
\begin{equation}
    \begin{split}
        & \frac{4}{3} e^{-\frac{\alpha}{2}}\Bigl[ 2i\hbar \Bigl( \partial_+ S_0 \partial_+ + \partial_- S_0 \partial_- - \partial_\alpha S_0 \partial_\alpha \Bigl) \\
        & \ \ \ \ \ \ \ \ \ \ \ + \frac{\hbar^2}{M} \left( \partial_+^2 + \partial_-^2 - \partial_\alpha^2 \right) \Bigr] \psi_2 \\
        & + \frac{1}{2} \sum_{\textbf{k},\lambda} \left[ \omega_{k,\eta}^2 (v_\textbf{k}^\lambda)^2 + \V - \hbar^2 \partial_{v_\textbf{k}^\lambda}^2 \right] \psi_2 = 0
        \label{5.gaugefix-1}
    \end{split}
\end{equation}
which can be recast into a differential equation for $\theta_2$. The lack of an order $M$ in the expansion \eqref{5.thetaexp} is due to the absence of a matter contribution in \eqref{5.HJ} at order $M$, namely the Planckian scale: at that order, the gauge transformation is simply not allowed without breaking the Born-Oppenheimer separation.

After condition \eqref{5.gaugefix-1} has been implemented, one deals with the much simplified form of \eqref{5.WDWM-1lunga}
\begin{equation}
    \begin{split}
        & \frac{1}{2} \sum_\textbf{k} \left[ \nu_{k,\eta}^2 \phi_\textbf{k}^2 - \hbar^2 \partial_{\phi_\textbf{k}}^2 \right] \chi_2 - i\hbar \ e^\alpha W^{-1} \frac{\delta}{\delta T} \chi_2 \\
        & + \frac{1}{2} \sum_{\textbf{k},\lambda} \left[ - \hbar^2 \partial_{v_\textbf{k}^\lambda}^2 - 2i\hbar \partial_{v_\textbf{k}^\lambda} S_1 \partial_{v_\textbf{k}^\lambda} \right] \chi_2 = 0,
    \end{split}
\end{equation}
where the gravitational wave functional $\psi_2$, acting as purely multiplicative, has been simplified. With the spatial integration and definition \eqref{5.time}, we obtain
\begin{equation}
    i\hbar \frac{\delta}{\delta \tau} \chi_2 = \hat{\mathcal{H}}^\phi \chi_2 + \hat{\mathcal{H}}^{QG} \chi_2,
    \label{5.corrSchr}
\end{equation}
where 
\begin{equation}\label{5.defHqg}
    \hat{\mathcal{H}}^{QG} \equiv \frac{1}{2} \int_\Sigma d^3x \sum_{\textbf{k},\lambda} \left[ - \hbar^2 \partial_{v_\textbf{k}^\lambda}^2 - 2i\hbar \partial_{v_\textbf{k}^\lambda} S_1 \partial_{v_\textbf{k}^\lambda} \right]
\end{equation}
is the operator containing the quantum-gravity induced corrections to the matter dynamics \eqref{5.Schr}. This effect is encoded in the dependence of $\chi_2$ on the $\vkl$ and also related to the form of $S_1$, which in turn is fixed by the $\order{M^0}$ gauge-fixing condition \eqref{5.gaugefix0} (i.e., the gravitational WDW equation): thus, the quantum corrections depend on the specific dynamics that gravitons follow on the classical background.

\subsection{QFT-CS phenomenology as \\ an effective theory}\label{ssec4:fluid-gravitons-effective}

The recovery of the functional Schr\"{o}dinger equations \eqref{5.Schr} and \eqref{5.corrSchr} for the matter sector does not imply the correct QFT limit of our system, since $\chi$ still depends on the cosmological perturbations $\vkl$ of the background (as discussed in Sec.~\ref{sec3:gravitons-vilenkin}). The idea is therefore to average over those degrees of freedom, in order to recover the standard QFT phenomenology as an effective theory.

At $\order{M^0}$, this procedure can be easily performed since all operators in \eqref{5.Schr} act on the matter variables alone, thus multiplying by $\psi_1^*\psi_1$ allows us to rewrite
\begin{equation}
    \hat{\mathcal{H}}^\phi \left( \psi_1^* \psi_1 \chi_1 \right) = i \hbar \frac{\delta}{\delta \tau} \left( \psi_1^* \psi_1 \chi_1 \right).
\end{equation}
For the same reason $\hat{\mathcal{H}}^\phi$ and $\delta/\delta \tau$ commute with the integration over the $\vkl$; thus, one can define an averaged wave functional
\begin{equation}
    \Theta_1(\phik,T;\alpha,\beta_\pm) = \int \prod_{\textbf{k},\lambda} d\vkl \ \psi_1^* \psi_1 \chi_1
    \label{6.Theta1}
\end{equation}
in clear analogy to \eqref{3.theta-averaged}. We stress that we have also a dependence on the reference fluid variable, which satisfies 
\begin{equation}
    \hat{\mathcal{H}}^\phi \Theta_1 = i \hbar \frac{\delta}{\delta \tau} \Theta_1.
    \label{6.avSchr}
\end{equation}
Since $\Theta_1$ depends only on the classical background, Eq.~\eqref{6.avSchr} clearly reproduces the phenomenology of QFT on curved spacetime. 

The $\order{M^{-1}}$ is less straightforward. We define the averaged wave functional
\begin{equation}
    \Theta_2(\phik,T;\alpha,\beta_\pm) = \int \prod_{\textbf{k},\lambda} d\vkl \ \psi_2^* \psi_2 \chi_2
\end{equation}
and wish to recast the dynamics \eqref{5.corrSchr} as acting on $\Theta_2$. However, while one can multiply Eq.~\eqref{5.corrSchr} by $\psi_2^*\psi_2$, this factor cannot be brought inside $\hat{\mathcal{H}}^{QG}$, since this operator contains derivatives with respect to the $\vkl$ and thus it does not commute with the averaging. In this sense, it seems that this averaging procedure cannot be implemented in the gauge corresponding to Eq.~\eqref{5.corrSchr}.
However, one can still recover QFT on average at this order by paying the price of altering the gauge-fixing condition \eqref{5.gaugefix-1}. In particular, by selecting the different gauge
\begin{equation}
    \begin{split}
        & \frac{4}{3} e^{-\frac{\alpha}{2}} \psi_2^* \Bigl[ 2i\hbar \left( \partial_+ S_0 \partial_+ + \partial_- S_0 \partial_- - \partial_\alpha S_0 \partial_\alpha \right) \\
        & \ \ \ \ \ \ \ \ \ \ \ + \frac{\hbar^2}{M} \left( \partial_+^2 + \partial_-^2 - \partial_\alpha^2 \right) \Bigr] \psi_2 \\
        & + \frac{1}{2} \sum_{\textbf{k},\lambda} \psi_2^* \left[ \omega_{k,\eta}^2 (v_\textbf{k}^\lambda)^2 + \V - \hbar^2 \partial_{v_\textbf{k}^\lambda}^2 \right] \psi_2 \\
        & + \frac{1}{2} \sum_{\textbf{k},\lambda} \left[ \hbar^2 \partial_\vkl^2 - 2i\hbar \partial_\vkl S_1 \partial_\vkl \right] \left( \psi_2^* \psi_2 \right) \\
        & + i\hbar \sum_{\textbf{k},\lambda} \psi_2^* (\partial_\vkl^2 S_1) \psi_2 = 0,
        \label{6.newgauge}
    \end{split}
\end{equation}
and integrating over the $\vkl$ with the following boundary conditions
\begin{equation}
    \begin{split}
        \int \prod_{\textbf{k},\lambda} d\vkl \sum_{\textbf{k},\lambda} & \hbar^2 \partial_\vkl \left[ \psi_2^* \psi_2 \partial_\vkl \chi_2 - \partial_\vkl \left( \psi_2^* \psi_2 \chi_2 \right) \right.\\
        &\left. + \frac{i}{\hbar} (\partial_\vkl S_1) \psi_2^* \psi_2 \chi_2 \right] = 0,
    \end{split}
    \label{6.boundary}
\end{equation}
the constraint equation at $\order{M^{-1}}$ brings
\begin{equation}
    i\hbar \frac{\delta}{\delta \tau} \Theta_2 = \hat{\mathcal{H}}^\phi \Theta_2.
    \label{6.avcorrSchr}
\end{equation}
This result represents a striking difference with the $\order{M^0}$ case: there the gauge-fixing condition \eqref{5.gaugefix-1} was chosen to represent the gravitational WDW equation, while here the interpretation of Eq.~\eqref{6.newgauge} seems more obscure. We also stress that such choice actually cancels out all quantum corrections due to the presence of gravitons.

If we wish to describe the quantum modifications due to the gravitons' modes $v_\textbf{k}^\lambda$ and at the same time to interpret the gauge-fixing condition as the gravitational WDW at all orders, then we must deal with Eq.~\eqref{5.corrSchr} for the nonaveraged wave functional $\chi_2$. The effective description of matter with small deviation due to the gravitons can be obtained by solving Eq.~\eqref{5.corrSchr}for $\chi_2$ and performing the integration over the $\vkl$ on this solution
\begin{equation}
    \int \prod_{\textbf{k},\lambda} d\vkl \, \chi_2\,,
\end{equation}
which does not coincide with the averaging of\eqref{5.corrSchr} over the $\vkl$. Clearly, this leaves the gauge choice \eqref{5.gaugefix-1} untouched, i.e., representing the gravitational constraint, and the matter dynamics will still present quantum corrections attributed to gravitons as mean-field effects, as we will show in Sec.~\ref{sec5:fluid-gravitons-spectrum}.

\section{Kucha\v{r}-Torre clock for the inflationary Universe}\label{sec5:fluid-gravitons-spectrum}

In this section we provide a simple cosmological application of the proposed model, i.e., the inflationary scenario with a single scalar field rolling in the region of approximately constant potential.

Let us first stress the implications of the inflation potential in such formulation: since $V(\phi)\simeq const.$ acts as a cosmological constant contribution, the minisuperspace background cannot be a vacuum geometry but it must contain $\Lambda$ as a source. Therefore, we must move our choice of the classical background from the Bianchi I of Sec. \ref{sec3:gravitons-vilenkin}, \ref{sec4:fluid-gravitons} to a different one, namely the Friedmann-Lemaitre-Robertson-Walker (FLRW) geometry. In such case, one should in principle deal with scalar perturbations of the metric along with tensor ones; however, the independence of the two allows us to consider once again only tensor contributions as ``slow'' gravity fluctuations in our model.
This means that we are neglecting the backreaction of the scalar field on the spacetime geometry, coherently with the Born-Oppenheimer separation previously implemented. 

The inflationary phase is identified with the de Sitter one by neglecting the slow-roll parameter $\varepsilon=-\dot{\Hubble}/\Hubble^2$ (being $\Hubble = \dot{a}/a$ the Hubble parameter), so that $V(\phi)$ reconstructs the cosmological constant term; more precisely, in order to avoid the singular behavior for $\varepsilon \rightarrow 0$, we shall fix the it to a small nonzero constant, as in \cite{bib:brizuela-kiefer-2016-desitter,bib:maniccia-montani-torcellini}. 

The FLRW line element in the conformal gauge reads:
\begin{equation}
    ds^2 = a^2(\eta) \Bigl[ d\eta^2 - dx^2 - dy^2 - dz^2 \Bigr],
    \label{7.FLRW}
\end{equation}
where $a(\eta)$ is the cosmic scale factor, whose solution is given by
\begin{equation}
    a(\eta)=-\frac{1}{\Hubble_0 \eta}
    \label{7.a}
\end{equation}
being $\Hubble_0$ the constant Hubble parameter. This sets up the classical background, over which we consider tensor gravitational fluctuations and then the inflaton field living on a such perturbed geometry (scalar perturbations are attributed to the inflaton only). 

Following the gauge-invariant formalism, both the graviton perturbations $\vkl$ and the inflaton modes $\phik$ take the form of time-dependent harmonic oscillators on the FLRW background with the same frequency, being $\epsilon = const$ \cite{bib:brizuela-kiefer-2016-desitter}
\begin{equation}
    \omega_{k,\eta}^2=\nu_{k,\eta}^2=k^2-\frac{2}{\eta^2}.
    \label{7.freq}
\end{equation}
However, the inflaton fluctuations present a distinct role as they are crucial in shaping the primordial structure of the Universe. When the physical wavelength of this fluctuations $\lambda_{phys} = a(\eta)\lambda_0$ ($\lambda_0 = $ comoving wavelength) is larger than the microphysics horizon, defined by the Hubble radius $\Hubble^{-1} = a/\dot{a}$, the evolution is dominated by gravity; while for $\lambda_{phys} \ll H^{-1}$ quantum matter prevails with minimal disturbance from curvature \cite{bib:rubakov-inflation}. Here, $\dot{a}$ stands for the derivative with respect to the synchronous time, in terms of which the scale factor grows exponentially: $a(t)=e^{\Hubble_0 t}$.
In a de Sitter background,the Hubble radius is constant, while $\lambda_{phys}$ increases exponentially. Therefore quantum fluctuations emerge at early times and rapidly grow, going outside the horizon and then reentering the Hubble radius after inflation when $\Hubble^{-1}$ grows faster than $\lambda_{phys}$; in other words, they take the form of density perturbations that ultimately lead to the formation of structures \cite{bib:MVP-2012}.

\subsection{Model setup up to $\order{M^{-1}}$}\label{ssec5:setup}
The full WDW constraint is now composed of an FLRW classical background, graviton perturbations, the scalar inflaton field and the reference fluid. The constant inflaton potential can be interpreted as a cosmological constant $\Lambda$ and thus is absorbed into the background contribution, so that $\varphi$ behaves like a free massless scalar field. Thus the super-Hamiltonian contributions of gravitons, inflaton, and reference fluid are similar to those we previously discussed, while the main difference lies in the background term describing the FLRW geometry:
\begin{equation}\label{7.Hdesitter}
    \hat{H}^{dS} = \frac{\hbar^2}{48M} e^{-3\alpha} \partial_\alpha^2 + 4M\Lambda e^{3\alpha},
\end{equation}
where $\alpha=\ln{a}$ is the Misner variable that describes this minisuperspace geometry. Note that due to isotropy the variables $\beta_\pm$ are forced to be identically zero, and also the term $\V$ mixing different polarization states in the gravitons sector is absent:
\begin{equation}
    \hat{H}^{v^\lambda} = \frac{e^{-\alpha}}{2} \sum_{\textbf{k},\lambda} \left[ -\hbar^2 \partial_\vkl^2 + \omega_{k,\eta}^2(\vkl)^2 \right].
\end{equation}
Implementing the Born-Oppenheimer separation and WKB expansion in $1/M$ of the wave functional $\Psi$
\begin{equation}
    \Psi = e^{\frac{i}{\hbar}(MS_0+S_1+\frac{1}{M}S_2)} e^{\frac{i}{\hbar}(Q_1+\frac{1}{M}Q_2)}\,,
    \label{7.BOWKB}
\end{equation}
we can recast the total constraint
\begin{equation}
    \Bigl( \hat{H}^{dS} + \hat{H}^{v^\lambda} + \hat{H}^\phi + \hat{H}^f \Bigr) \Psi = 0
\end{equation}
as a hierarchy of equations. Clearly, the $\order{M^2}$ states that $S_0$ is independent of the $\vkl$, namely $S_0=S_0(\alpha)$; the $\order{M}$ one reconstructs the Hamilton-Jacobi equation of the classical background:
\begin{equation}
    \frac{1}{48}e^{-3\alpha}(\partial_\alpha S_0)^2 - 4\Lambda e^{3\alpha} = 0\,,
\end{equation}
which can be solved by separation of variables:
\begin{equation}
    S_0(\alpha) = - 8 \sqrt{\frac{\Lambda}{3}} \left( e^{3\alpha} - e^{3\alpha_0} \right)\,,
    \label{7.S0}
\end{equation}
where we have chosen the negative solution to reproduce an expanding Universe \textcolor{blue}{[$p_{\alpha}<0$, with the same reasoning of Eq.~\eqref{3.S0}]}.

At $\order{M^0}$ we use the gauge freedom of the Born-Oppenheimer separation to impose the gravitational constraint separately, i.e.,
\begin{equation}
    \begin{split}
        & \ \ \ \frac{e^{-2\alpha}}{24}  \left( i\hbar \partial_\alpha^2 S_0 -2 \partial_\alpha S_0 \partial_\alpha S_1 \right) \\
        & + \sum_{\textbf{k},\lambda} \left[ -i\hbar \partial_{v_\textbf{k}^\lambda}^2 S_1 + (\partial_{v_\textbf{k}^\lambda} S_1)^2 + \omega_{k,\eta}^2(v_\textbf{k}^\lambda)^2 \right] = 0,
    \end{split}
\end{equation}
which in terms of $\psi_1 = \exp(iS_1/\hbar)$ reads
\begin{equation}
    \begin{split}
        & \frac{e^{-2\alpha}}{24} \psi_1^* \left( i\hbar \partial_\alpha^2 S_0 + 2i\hbar \partial_\alpha S_0 \partial_\alpha \right) \psi_1 \\
        & + \psi_1^* \sum_{\textbf{k},\lambda} \left[ -\hbar^2 \partial_{v_\textbf{k}^\lambda}^2 + \omega_{k,\eta}^2(v_\textbf{k}^\lambda)^2 \right] \psi_1 = 0.
    \end{split}
\label{7.gaugefixing}
\end{equation}
Now the constraint equation for $\chi_1=\exp(iQ_1/\hbar)$
\begin{equation}\label{7.schrodinger}
    i\hbar \ e^{\alpha} W^{-1} \frac{\delta}{\delta T} \chi_1 = \frac{1}{2} \sum_\textbf{k} \left[ - \hbar^2 \partial_{\phi_\textbf{k}}^2 + \nu_{k,\eta}^2 \phi_\textbf{k}^2 \right] \chi_1
\end{equation}
can be recast by defining the clock as in \eqref{5.time}, giving the same Schr\"{o}dinger dynamics of \eqref{5.Schr}. We stress that the gauge-fixing condition \eqref{7.gaugefixing} is different from \eqref{5.gaugefix0} since we here consider an FLRW geometry, and the same holds for the frequencies $\nu_{k,\eta}^2$ and $\omega_{k,\eta}^2$, see \eqref{7.freq}; this means that the solutions $\psi_1$ and $\chi_1$ are different from the Bianchi I case. However, since the Schr\"{o}dinger functional dynamics for $\chi_1$ is identical, we can perform the same averaging procedure of Sec.~\ref{sec4:fluid-gravitons} defining $\Theta_1$ as in \eqref{6.Theta1} which again satisfies \eqref{6.avSchr}. Such equation represents the desired QFT limit up to $\order{M^0}$, having removed the dependence on the graviton fluctuations.

The $\order{M^{-1}}$ equation
\begin{equation}
    \begin{split}
        & \frac{e^{-3\alpha}}{48}  \Bigl[ i\hbar \partial_\alpha^2 S_1 - (\partial_\alpha S_1 )^2 - 2 \partial_\alpha S_0 \partial_\alpha S_1 \Bigr] \\
        & + \frac{1}{2} \sum_{\textbf{k},\lambda} \Bigl[ - i\hbar \partial_{v_\textbf{k}^\lambda}^2 + 2 \partial_{v_\textbf{k}^\lambda} S_1 \partial_{v_\textbf{k}^\lambda} \Bigr] \Bigl( S_2 + MQ_1 \Bigr) \\
        & + \frac{1}{2} \sum_\textbf{k} \Bigl[ - i\hbar \partial_{\phi_\textbf{k}}^2 Q_2 + 2 \partial_{\phi_\textbf{k}} Q_1 \partial_{\phi_\textbf{k}} Q_2 \Bigr] \\
        & + e^\alpha W^{-1} \frac{\delta Q_2}{\delta T} = 0
    \end{split}
\end{equation}
combined with the $\order{M^0}$ one gives, for the wave functionals $\psi_2$ and $\chi_2$ [defined in \eqref{5.def-psi2-chi2}]
\begin{widetext}
\begin{equation}
    \begin{split}
        & \Biggl( \frac{e^{-3\alpha}}{48} \Biggl[ i\hbar \partial_\alpha S_0 \partial_\alpha + \frac{\hbar^2}{M} \partial_\alpha^2 \Biggr] \psi_2+ \frac{1}{2} \sum_{\textbf{k},\lambda} \Bigl[ \omega_{k,\eta}^2 (v_\textbf{k}^\lambda)^2 - \hbar^2 \partial_{v_\textbf{k}^\lambda}^2 \Bigr] \psi_2 \Biggr) \chi_2 \\
        & + \psi_2 \Biggl( \frac{1}{2} \sum_\textbf{k} \Bigl[ \nu_{k,\eta}^2 \phi_\textbf{k}^2 - \hbar^2 \partial_{\phi_\textbf{k}}^2 \Bigl] \chi_2  -i\hbar \ e^\alpha W^{-1} \frac{\delta}{\delta T} \chi_2 + \frac{1}{2} \sum_{\textbf{k},\lambda} \Bigl[ - \hbar^2 \partial_{v_\textbf{k}^\lambda}^2 - 2i\hbar \partial_{v_\textbf{k}^\lambda} S_1 \partial_{v_\textbf{k}^\lambda} \Bigr] \chi_2 \Biggr) = 0.
    \end{split}
    \label{7.WDW-1}
\end{equation}
\end{widetext}
The first line, which is easily recognised as the gravitational constraint, can be removed using the $\order{M^{-1}}$ gauge freedom to impose the following condition on $\psi_2$:
\begin{equation}
    \begin{split}
        & \frac{e^{-3\alpha}}{24} \psi_2^* \left[ i\hbar \partial_\alpha S_0 \partial_\alpha + \frac{\hbar^2}{M} \partial_\alpha^2 \right] \psi_2 \\
        & + \psi_2^* \sum_{\textbf{k},\lambda} \Bigl[ \omega_{k,\eta}^2 (v_\textbf{k}^\lambda)^2 - \hbar^2 \partial_{v_\textbf{k}^\lambda}^2 \Bigr] \psi_2 = 0.
    \end{split}
\end{equation}
As in the previous section, integration on the hypersurface $\Sigma$ and the clock definition \eqref{5.time} bring Eq.~\eqref{7.WDW-1} in the form of a corrected Schr\"{o}dinger equation \eqref{5.corrSchr}, with the operator \eqref{5.defHqg} encoding the quantum corrections. We stress that both the time parameter and the operator $\hat{\mathcal{H}}^{QG}$ take the same form of the Bianchi I case, while the specific gravitational solutions $S_n$ are now different.

Now we cannot simply average Eq.~\eqref{5.corrSchr} on the gravitons' degrees of freedom $v_\textbf{k}^\lambda$ in order to describe the dynamics of the averaged wave functional $\Theta_2$, but we would have to modify the choice of the gauge making its physical interpretation less straightforward (see discussion in Sec.~\ref{ssec4:fluid-gravitons-effective}). Therefore, we will first solve the dynamics of $\chi_2$ and then perform the averaging procedure on the solution itself.

\subsection{Gravitons sector}\label{ssec5:graviton-dynamics}

In order to specify the form of the operator \eqref{5.defHqg} representing quantum gravity corrections in \eqref{5.corrSchr}, we must first solve the gauge-fixing condition \eqref{7.gaugefixing} for $\psi_1$ and compute
\begin{equation}
    S_1(\alpha,\vkl)=-i\hbar \ln \psi_1(\alpha,\vkl)\,.
    \label{7.S1ln}
\end{equation}
We recall that Eq.~\eqref{7.gaugefixing} expresses the dynamics of gravitons on the classical background, however it does not include the reference fluid clock i.e. it represents a timeless dynamics. The gravitons' time evolution can be recovered by exploiting the classical evolution of the Misner variable $\alpha = \alpha(\eta)$, described by the Hamilton equation
\begin{equation}
    \dot{\alpha} = N \frac{\partial H^{dS}}{\partial p_\alpha},
\end{equation}
where $H^{dS}$ is the classical version of \eqref{7.Hdesitter}. Therefore, in the conformal gauge $N=e^\alpha$ 
\begin{equation}
    \dot{\alpha} = - \frac{p_\alpha}{24M} e^{-2\alpha} \longrightarrow p_\alpha = - 24 M e^{2\alpha} \dot{\alpha}\,.
\end{equation}
Since $M S_0$ is the classical action reproducing the gravitational background, by definition $p_\alpha = M \partial_\alpha S_0$ so that
\begin{equation}
    \frac{\partial S_0}{\partial \alpha} = -24 e^{2\alpha} \dot{\alpha}
\end{equation}
which implies 
\begin{equation}
    \frac{\partial S_0}{\partial \alpha} \frac{\partial}{\partial \alpha} = - 24 e^{2\alpha} \dot{\alpha} \frac{\partial}{\partial \alpha} = - 24 e^{2\alpha} \frac{\partial}{\partial \eta}
    \label{7.deta}
\end{equation}
due to the fact that $\dot{\alpha}=d\alpha/d\eta$ in this gauge. Moreover, from the explicit solution \eqref{7.S0} we have
\begin{equation}
    \frac{\partial^2 S_0}{\partial \alpha^2} = - 24 \sqrt{3\Lambda} \ e^{3\alpha}.
    \label{7.d2S0}
\end{equation}
Now making use of Eqs.~\eqref{7.deta}-\eqref{7.d2S0}, the gravitational constraint \eqref{7.gaugefixing} can be recast as
\begin{equation}
    \begin{split}
        i\hbar \Bigl( & \sqrt{3\Lambda} e^\alpha + 2 \partial_\eta \Bigr) \psi_1 \\
        & = \sum_{\textbf{k},\lambda} \left[ -\hbar^2 \partial_\vkl^2 + \omega_{k,\eta}^2(\vkl)^2 \right] \psi_1,
    \end{split}
    \label{7.etagravWDW}
\end{equation}
which represents the gravitons' dynamics in the conformal time $\eta$. 
Let us clarify this statement: the gravitons' evolution is parametrized by the (classical) conformal time $\eta$, while the matter sector is parametrized by the fluid time $\tau$. However, the diffeomorphism invariance of the theory allows us to remove such ambiguity. In fact, one can perform a coordinate transformation so that, at each spacetime point $x$, the label time $t$ corresponds exactly to the Gaussian time (see also \cite{bib:KucharTorre}):
\begin{equation}
T(x) = T(t, \textbf{x}) = t\,.
\end{equation}
Indeed, among all possible coordinate choices to express Eq.~\eqref{7.gaugefixing}, we take the one for which the time variable is identical to the Gaussian one by definition of $T$, so that the label time parametrizing the gravitons coincides with the reference fluid time. 
Clearly, this identification can only be performed after deriving the functional Schr\"odinger equation and gauge-fixing condition \eqref{7.gaugefixing}, so that $T$ is still treated as a quantum variable and the separation \eqref{7.BOWKB} is still meaningful. 
Moreover, in the synchronous gauge $N=1$ with $T=t$ we have $W^{-1}(x)=1$ identically, so that the fluid time can be directly related to $\eta$:
\begin{equation}
    \frac{\delta}{\delta T} = \frac{\partial}{\partial t} = \frac{1}{N} \frac{\partial}{\partial \eta} = e^{-\alpha} \frac{\partial}{\partial \eta}\,,
\end{equation}
and the factor $e^\alpha$ in \eqref{5.time} is simplified, giving
\begin{equation}
    \frac{\delta}{\delta \tau} = \int_\Sigma d^3x \ \frac{\partial}{\partial \eta}.
    \label{7.taueta}
\end{equation}
This correspondence between $t$ and $T$, seen as exploiting the diffeomorphism invariance of the theory, allows us to recover a description in which all quantum degrees of freedom evolve with the same time parameter, while still preserving the difference between the classical variable $\eta$ and the quantum one $T$.

Now we are still left with solving the $\order{M^0}$ gravitational constraint, recast in the form \eqref{7.etagravWDW}. First, we observe that by performing the following transformation
\begin{equation}
    \psi_1 \rightarrow \bar{\psi}_1 = \psi_1 e^{f(\eta)}
\end{equation}
with $f(\eta)$ such that
\begin{equation}
    f'(\eta) = - \frac{\sqrt{3 \Lambda}}{2} e^{\alpha(\eta)}\,,
\end{equation}
we have that the cosmological constant term in \eqref{5.time} is reabsorbed, giving the following simplified expression for $\bar{\psi}_1$
\begin{equation}
    i\hbar \partial_\eta \bar{\psi}_1 = \sum_{\textbf{k},\lambda} \left[ -\hbar^2 \partial_\vkl^2 + \omega_{k,\eta}^2(\vkl)^2 \right] \bar{\psi}_1\,,
    \label{7.timedepHO}
\end{equation}
i.e. we have mapped the problem to the standard time-dependent harmonic oscillator. 
Recalling that $e^{\alpha(\eta)}=a(\eta)$ and the classical solution \eqref{7.a} in the FLRW geometry, $f(\eta)$ can be specified as
\begin{equation}
    f(\eta) = - \frac{\sqrt{3\Lambda}}{2} \int d\eta \ a(\eta) = \frac{\sqrt{3\Lambda}}{2H_0} \ln (-\eta)\,,
\end{equation}
where the integration constant has been taken to unit for simplicity. Then the transformation between $\bar{\psi}_1$ and $\psi_1$ is
\begin{equation}
    \bar{\psi}_1 = (-\eta)^{\frac{\sqrt{3\Lambda}}{2H_0}} \psi_1\,,
    \label{7.psi'}
\end{equation}
so that we can solve Eq.~\eqref{7.timedepHO} for $\bar{\psi}_1$ and then obtain the gravitational wave functional $\psi_1$ by inverting \eqref{7.psi'}.

Since Eq.~\eqref{7.timedepHO} describes the dynamics of time-dependent harmonic oscillator, we can solve it by means of the Lewis-Riesenfeld invariant method \cite{bib:LR1,bib:LR2,bib:LR3}. We also implement the Bunch-Davies vacuum requirement: the ground state of the physical wave functional must correspond to the Minkowski vacuum in the limit $\eta \rightarrow -\infty$, i.e. when the gravitons' wavelength is much smaller than the curvature of the Universe, thus fixing the boundary condition. 
As a matter of fact, this is formally identical to the treatment \cite{bib:maniccia-montani-torcellini} for the matter wave functional in the absence of graviton perturbations. 
With the same procedure, we find the solution for $\bar{\psi}_1$ as
\begin{widetext}
\begin{equation}
    \begin{split}
        \bar{\psi}_1 (\eta, v_\textbf{k}^\lambda) = \prod_{\textbf{k},\lambda} \left[\frac{k^3}{\pi \hbar \left( \frac{1}{\eta^2} + k^2 \right)} \right]^{\frac{1}{4}}  \exp \left[ - \frac{i}{2} \Bigl(\eta k - \arctan(\eta k) \Bigr) \right] \exp \left[ \frac{i}{2\hbar} \frac{ik^3\eta^3-1}{\eta^3 \left(\frac{1}{\eta^2} + k^2 \right)} (v_\textbf{k}^\lambda)^2 \right]
    \end{split}
\end{equation}
\end{widetext}
which, inverting Eq.~\eqref{7.psi'} and using \eqref{7.S1ln}, gives the following solution for $S_1$:
\begin{widetext}
\begin{equation}
    \begin{split}
        S_1 (\eta,v_\textbf{k}^\lambda) = - i \hbar \sum_{\textbf{k},\lambda} \left[ \frac{1}{4} \ln \left(\frac{k^3}{\pi \hbar \left( \frac{1}{\eta^2} + k^2 \right)} \right)  - \frac{\sqrt{3\Lambda}}{2\Hubble_0} \ln (-\eta) - \frac{i}{2} \Bigl( \eta k - \arctan(\eta k) \Bigr) + \frac{i}{2\hbar} \frac{ik^3\eta^3-1}{\eta^3 \left(\frac{1}{\eta^2} + k^2 \right)}  (v_\textbf{k}^\lambda)^2 \right]
    \end{split}
\end{equation}
\end{widetext}
We are interested in the quantity $\partial_\vkl S_1$ that enters the quantum gravity corrections, which can be written as
\begin{equation}
    \partial_\vkl S_1 = \frac{i}{\hbar} \xi_k(\eta) \vkl
    \label{7.dS1dv}
\end{equation}
having defined the function
\begin{equation}
    \xi_k(\eta) = \frac{ik^3\eta^3-1}{\eta^3 \left(\frac{1}{\eta^2} + k^2 \right)}.
    \label{7.xi}
\end{equation}
The linearity of $\partial_\vkl S_1$ in the $\vkl$, which enters the quantum corrections operator $\hat{\mathcal{H}}^{QG}$, will play an important role in determining the matter sector solution.

\subsection{Inflaton solution with Gaussian ansatz}\label{ssec5:gaussian-asatz}

We now focus on the matter dynamics expressed by \eqref{5.corrSchr} which, apart from an integration over the hypersurface $\Sigma$, takes the form 
\begin{equation}
    \begin{split}
        i \hbar \partial_\eta \chi_2 = & \frac{1}{2} \sum_\textbf{k} \Bigl[ - \hbar^2 \partial_\phik^2 + \nu_{k,\eta}^2 (\phik)^2 \Bigr] \chi_2 \\
        & - \frac{1}{2} \sum_{\textbf{k},\lambda} \Bigl[ \hbar^2 \partial_\vkl^2 + 2 i \hbar \partial_\vkl S_1 \partial_\vkl \Bigr] \chi_2.
        \label{7.mattereq}
    \end{split}
\end{equation}
From now on, in comparison with existing literature, we shall use units in which $\hbar=1$ and we drop the subscript in $\chi_2$, labeling it as $\chi$, to improve readability. 

Let us write the matter wave functional as a product of independent modes:
\begin{equation}
    \chi = \prod_\textbf{k} \chi_\textbf{k}
\end{equation}
where each mode depends on the respective variables $\phik$, $v_\textbf{k}^+$ and $v_\textbf{k}^\times$ for fixed values of $\textbf{k}$. This ansatz is based on the underlying assumption that both polarization states $+$ and $\times$ contribute the same amount to the total functional. Eq.~\eqref{7.mattereq} is then decomposed into a set of independent mode equations:
\begin{equation}
    \begin{split}
        i \partial_\eta \chi_\textbf{k} = & \frac{1}{2} \Bigl[ - \partial_\phik^2 + \nu_{k,\eta}^2 (\phik)^2 \Bigr] \chi_\textbf{k} \\
        & - \frac{1}{2} \sum_\lambda \Bigl[ \partial_\vkl^2 + 2 i \partial_\vkl S_1 \partial_\vkl \Bigr] \chi_\textbf{k}.
        \label{7.modeeq}
    \end{split}
\end{equation}

Following the lines of \cite{bib:brizuela-kiefer-2016-desitter,bib:maniccia-montani-torcellini}, we shall look for a solution of \eqref{7.modeeq} in Gaussian form:
\begin{equation}
    \chi_\textbf{k} = A_\textbf{k}(\eta,v_\textbf{k}) \ e^{-\frac{1}{2} \Omega_\textbf{k}(\eta) (\phik)^2}
    \label{7.Gauss}
\end{equation}
where $v_\textbf{k}$ implies both polarizations $(v_\textbf{k}^+,v_\textbf{k}^\times)$. We let $A_\textbf{k}$ be a complex amplitude, so it can be written in terms of its (real) magnitude $\mathcal{N}_\textbf{k}$ and phase $\Upsilon_\textbf{k}$ as
\begin{equation}
    A_\textbf{k}(\eta,v_\textbf{k}) = \mathcal{N}_\textbf{k}(\eta,v_\textbf{k}) \ e^{i \Upsilon_\textbf{k}(\eta,v_\textbf{k})}\,,
\end{equation}
and the Gaussian width $\Omega_\textbf{k}$ depends on $\eta$ alone. 
In the spirit of searching for a WKB expanded solution of the matter sector, we can write
\begin{gather}
    \mathcal{N}_\textbf{k}(\eta,v_\textbf{k}) = N_\textbf{k}(\eta) \left[ 1 + \frac{\Hubble_0^2}{M}G_\textbf{k}(\eta, v_\textbf{k}) \right], \label{7.N} \\
    \ \Upsilon_\textbf{k}(\eta,v_\textbf{k}) = \Upsilon_\textbf{k}^{(0)}(\eta) + \frac{\Hubble_0^2}{M} \Upsilon_\textbf{k}^{(1)}(\eta,v_\textbf{k}), \label{7.phi} \\
    \Omega_\textbf{k} (\eta) = \Omega_\textbf{k}^{(0)}(\eta) + \frac{\Hubble_0^2}{M} \Omega_\textbf{k}^{(1)}(\eta)\,,\label{7.Omega}
\end{gather}
where we have introduced the adimensional factor $\Hubble_0^2/{M}$ and exploited the property that effects from tensor perturbations appear only at $\order{M^{-1}}$, see Eqs.~\eqref{7.S0} and \eqref{7.schrodinger}, so that all $\order{M^0}$ contributions depend on $\eta$ alone. For an analogous expansion of the Gaussian ansatz, see \cite{bib:brizuela-kiefer-2016-desitter}.
Therefore we work with the mode functional $\chi_\textbf{k}$
\begin{equation}
    \begin{split}
        \chi_\textbf{k} = & N_\textbf{k}(\eta) \left( 1 + \frac{\Hubble_0^2}{M} G_\textbf{k}(\eta,v_\textbf{k}) \right) \\
        & \cdot \exp \left[ i \left( \Upsilon_\textbf{k}^{(0)}(\eta) + \frac{\Hubble_0^2}{M} \Upsilon_\textbf{k}^{(1)}(\eta,v_\textbf{k}) \right) \right] \\
        & \cdot \exp \left[ - \frac{1}{2} \left( \Omega_\textbf{k}^{(0)}(\eta) + \frac{\Hubble_0^2}{M}\Omega_\textbf{k}^{(1)}(\eta) \right) (\phik)^2 \right]
    \end{split}
    \label{7.newGauss}
\end{equation}
which, substituted into Eq.~\eqref{7.modeeq}, produces at $\order{M^0}$ and $\order{M^{-1}}$ the following equations
\begin{gather}
     i \partial_\eta \Omega_\textbf{k}^{(0)} = \left( \Omega_\textbf{k}^{(0)} \right)^2 - \nu_{k,\eta}^2, 
    \label{7.newOmega0eq} \\
    i \partial_\eta \Omega_\textbf{k}^{(1)} = 2 \Omega_\textbf{k}^{(0)} \Omega_\textbf{k}^{(1)}, \label{7.newOmega1eq} \\
    i \partial_\eta N_\textbf{k} - N_\textbf{k} \partial_\eta \Upsilon_\textbf{k}^{(0)} = \frac{1}{2} N_\textbf{k} \Omega_\textbf{k}^{(0)}, \label{7.newphi0eq} \\
    \begin{split}
        i & G_\textbf{k} \partial_\eta N_\textbf{k} + N_\textbf{k} \Bigl[ i \partial_\eta G_\textbf{k} - \partial_\eta \Upsilon_\textbf{k}^{(1)} - G_\textbf{k} \partial_\eta \Upsilon_\textbf{k}^{(0)} \Bigr] \\
        & = \frac{1}{2} N_\textbf{k} \Bigl[ G_\textbf{k} \Omega_\textbf{k}^{(0)} + \Omega_\textbf{k}^{(1)} - \sum_\lambda \Bigl( \partial_\vkl^2 G_\textbf{k} + i \partial_\vkl^2 \Upsilon_\textbf{k}^{(1)} \\
        & \ \ - 2 \xi_k(\eta) \left[ \partial_\vkl G_\textbf{k} + i \partial_{v_\vkl} \Upsilon_\textbf{k}^{(1)} \right] \vkl \Bigr) \Bigr], \label{7.newGeq}
    \end{split}
\end{gather}
where we expressed $\partial_\vkl S_1$ by means of \eqref{7.dS1dv}.

Equations~\eqref{7.newOmega0eq}--\eqref{7.newOmega1eq} can be immediately solved recalling the expression \eqref{7.freq} for $\nu_{k,\eta}^2$ in a de Sitter geometry; in particular, Eq.~\eqref{7.newOmega0eq} is identical to the one found in \cite{bib:brizuela-kiefer-2016-desitter} for the Gaussian width. By imposing the Bunch-Davies vacuum as the initial state, one finds 
\begin{equation}
    \Omega_\textbf{k}^{(0)}(\eta) = \frac{k^3\eta^2}{1+k^2\eta^2} + \frac{i}{\eta(1+k^2\eta^2)}.
\end{equation}
As previously discussed, all quantities that contribute to the perturbation spectrum of the inflaton field can be computed in the super-Hubble limit by taking $k \eta \rightarrow 0^-$, which gives
\begin{equation}
    \Omega_\textbf{k}^{(0)}(\eta) \sim k^3 \eta^2 + \frac{i}{\eta}.
    \label{7.Omega0}
\end{equation}
By substituting this expression into Eq.~\eqref{7.newOmega1eq}, the next-order Gaussian width results in
\begin{equation}
    \Omega_\textbf{k}^{(1)}(\eta) \sim c_1 \eta^2 \left( 1 - \frac{2}{3} i k^3 \eta^3 \right)\,,
    \label{7.Omega1}
\end{equation}
where $c_1$ is an integration constant.

We recall that the ansatz \eqref{7.newGauss} must be properly normalized over the $\phik$ and $\vkl$ modes:
\begin{equation}
    \int d\phik dv_\textbf{k}^+ dv_\textbf{k}^\times \ \chi_\textbf{k}^* \chi_\textbf{k} = 1\,,
\end{equation}
which gives a constraint on the solutions of the remaining equations:
\begin{equation}
    \begin{split}
        |N_\textbf{k}|^4 &  \left( 1 + \int dv_\textbf{k}^+ dv_\textbf{k}^\times \frac{4\Hubble_0^2}{M} G_\textbf{k} \right) \\
        & = \frac{\Re \Omega_\textbf{k}^{(0)}}{\pi} + \frac{\Hubble_0^2}{M}\frac{\Re \Omega_\textbf{k}^{(1)}}{\pi},
        \label{7.normcond}
    \end{split}
\end{equation}
where we neglected higher-order terms.
Here the presence of the fourth power in $N_\textbf{k}$ is due to the fact that, in principle, one needs to quantize separately the real and imaginary parts of the scalar perturbation variables, which are complex \cite{bib:MVP-2012}; however, dealing with $|\phik|^2$ instead of $(\phik)^2$ leads, although with a less rigorous formalism, to the same results \cite{bib:brizuela-kiefer-2016-desitter}.

Clearly, we are normalizing separately the terms of order $M^0$ and $M^{-1}$, since the former are independent of gravitons, so we limit the integration over $\vkl$ only on the latter [see \eqref{7.normcond}], finding
 \begin{gather}
    N_\textbf{k}(\eta) = \left( \frac{\Re \Omega_\textbf{k}^{(0)}(\eta)}{\pi} \right)^{\frac{1}{4}},  \label{7.Nnorm} \\
    \int dv_\textbf{k}^+ dv_\textbf{k}^\times \ G_\textbf{k}(\eta,v_\textbf{k}) = \frac{\Re \Omega_\textbf{k}^{(1)}(\eta)}{4 \Re \Omega_\textbf{k}^{(0)}(\eta)}. \label{7.Gnorm} 
\end{gather}
As we will show in the next Section, Eqs.~\eqref{7.newOmega0eq}--\eqref{7.newOmega1eq} --- along with the constraints \eqref{7.Nnorm}--\eqref{7.Gnorm} --- are enough to compute the perturbation spectrum we are interested in. The consistency of the Gaussian ansatz \eqref{7.newGauss} is shown in the Appendix.

\subsection{Power spectrum morphology}\label{ssec5:power-spectrum}

The computation of the inflaton perturbations spectrum relies on the corresponding two-point correlation function \cite{bib:weinberg-grav-cosm}:
\begin{equation}
    \begin{split}
        \Xi(\textbf{r}) & = \bra{0} \phi(\eta,\textbf{x})\phi(\eta,\textbf{x}+\textbf{r})\ket{0} \\
        & = \frac{1}{(2\pi)^3} \int d\textbf{k} \ e^{-i \textbf{k} \cdot \textbf{r}} \abs{f_\textbf{k}}^2,
        \label{7.corrfunc}
    \end{split}
\end{equation}
where $\ket{0}$ identifies the Bunch-Davies vacuum \cite{bib:BirrelDavies} and $f_\textbf{k}$ is the mode function associated to each Fourier mode $\phik$. Note that this expression retains a dependence on the conformal time $\eta$, which will be removed by considering the suitable limit. The power spectrum of such fluctuations is then obtained from the square of the Fourier amplitude in \eqref{7.corrfunc} as \cite{bib:PrimCosm,bib:MVP-2012}
\begin{equation}
    P_\phi(k) = \frac{k^3}{2\pi^2} \abs{f_\textbf{k}}^2\,.
\end{equation}
However, the scalar fluctuations $\phi$ can be related to the comoving curvature perturbation $\zeta$ using 
\begin{equation}
\zeta_\textbf{k} = \sqrt{\frac{4 \pi G}{\varepsilon}} \frac{\phi_\textbf{k}}{a}
\label{7.zeta}
\end{equation}
in Fourier space, with constant $\varepsilon$; this allows us to derive also the curvature perturbations spectrum:
\begin{equation}
P_\zeta(k) = \frac{2 G k^3}{\pi \varepsilon a^2} |f_\textbf{k}|^2. 
\label{7.Pzeta}
\end{equation}
We stress that such quantity must be computed in the super-Hubble limit $k/(aH) \ll 1$, namely $\lambda_{phys} \ll H^{-1}$, by taking $k\eta \rightarrow 0^-$. However, it can be shown that $\zeta_\textbf{k}$ is constant for all times in which the mode perturbation is outside the horizon \cite{bib:MVP-2012}; this allows to compute $P_\zeta(k)$ at the end of inflation, when the corresponding perturbations reenter the horizon.

Taking the Gaussian solution \eqref{7.newGauss} for $\chi_{\textbf{k}}$ as the vacuum state, the correlation function becomes
\begin{widetext}
\begin{equation}\label{7.corrfunction-gauss}
    \begin{split}
        \Xi(\textbf{r})  & = \int \prod_\textbf{p} d\phi_\textbf{p} \int \prod_\textbf{q} dv_\textbf{q}^+ dv_\textbf{q}^\times \left( \prod_\textbf{k} \chi_\textbf{k}^* \right) \phi(\eta,\textbf{x})  \phi(\eta,\textbf{x}+\textbf{r}) \left( \prod_{\textbf{k}'} \chi_{\textbf{k}'} \right) \\
        & = \prod_\textbf{k} |N_\textbf{k}|^4 \int dv_\textbf{k}^+ dv_\textbf{k}^\times \left( 1 + \frac{4}{M} G_\textbf{k} \right)  \int \prod_\textbf{p} d\phi_\textbf{p} \ e^{- \sum_{\textbf{k}'} \left[ \Re \Omega_{\textbf{k}'}^{(0)} + \frac{\Hubble_0^2}{M}\Re \Omega_{\textbf{k}'}^{(1)} \right] (\phi_{\textbf{k}'})^2} \phi(\eta,\textbf{x}) \phi(\eta,\textbf{x}+\textbf{r})
    \end{split}
\end{equation}
\end{widetext}
having carefully rearranged some factors and neglected all terms of $\order{M^{-2}}$ or higher. We recall that the integration in the $\vkl$ is restricted to terms $\propto M^{-1}$, as in \eqref{7.normcond}. This represents an averaging procedure of the matter wave functional \eqref{7.newGauss} over gravitons' degrees of freedom, instead of an averaged Eq.~\eqref{7.mattereq}, see discussion in Sec.~\ref{ssec4:fluid-gravitons-effective}, which does not modify the gauge fixing, i.e., the gravitational WDW equation still holds.

Equation~\eqref{7.corrfunction-gauss} differs from \cite{bib:brizuela-kiefer-2016-desitter,bib:maniccia-montani-torcellini} only in the presence of $\order{M^{-1}}$ contributions and integration over gravitons' degrees of freedom. Proceeding in a similar fashion, we use the normalization condition \eqref{7.normcond} and perform the 3D Fourier transform on $\phi$, decomposing the complex measure $d\phi_\textbf{p}$ into its real and imaginary parts $d\phi_\textbf{p}^R$ and $d\phi_\textbf{p}^I$ to compute the Gaussian integrals (we refer to \cite{bib:maniccia-montani-torcellini} for a more detailed discussion). We then obtain the following expression for the correlation function:
\begin{equation}
    \Xi(\textbf{r}) = \int \frac{d\textbf{p}}{(2\pi)^3} e^{-i \textbf{p} \cdot \textbf{r}} \frac{1}{2 \left[ \Re \Omega_\textbf{p}^{(0)} + \frac{\Hubble_0^2}{M}\Re \Omega_\textbf{p}^{(1)} \right]}
\end{equation}
which, compared to \eqref{7.corrfunc}, allows us to determine the quantity $|f_\textbf{k}|^2$, resulting in the following power spectrum for curvature perturbations \eqref{7.Pzeta}
\begin{equation}
    P_\zeta(k) = \frac{G k^3}{\pi \varepsilon a^2} \frac{1}{k^3 \eta^2 + \frac{\Hubble_0^2}{M} c_1\eta^2} \,.
\end{equation}
Here we have inserted the solutions \eqref{7.Omega0} and \eqref{7.Omega1} computed in the super-Hubble limit. Now recalling the scale factor behavior \eqref{7.a} and expanding in $1/M$, we obtain
\begin{equation}\label{eq:spettro-fin-marco}
    P_\zeta(k) = \frac{G \Hubble_0^2}{\pi \varepsilon} \left[ 1 - c_1 \frac{\Hubble_0^2}{M}\left(\frac{k_0}{k} \right)^3 \right]_{k=aH_0}
\end{equation}
evaluated at the horizon crossing after the end of inflation, as previously motivated. We have put in evidence a reference wave number $k_0$, coming from the discretized treatment of the perturbations as time-dependent harmonic oscillators, which recovers the correct dimensions, see \cite{bib:brizuela-kiefer-2016-desitter}.

Equation~\eqref{eq:spettro-fin-marco} presents a small $\order{M^{-1}}$ correction to the standard de Sitter spectrum obtained in \cite{bib:maniccia-montani-torcellini}, with a scale dependence $\propto k^{-3}$; this is in accordance with the results of \cite{bib:casadio-2006-spettro,bib:venturi-2013-spettro,bib:brizuela-kiefer-2016-desitter} computed in the de Sitter phase, \cite{bib:venturi-2014-spettro-slowroll,bib:venturi-2015-spettro-confronto,bib:brizuela-kiefer-2016-slow-roll} in the slow-roll approximation, and \cite{bib:venturi-2016-spettro} for power-law inflation, all with a time implementation analogous to \cite{bib:kiefer-1991} and without the gravitons treatment here presented. 
However, we stress that in \eqref{eq:spettro-fin-marco} the integration constant $c_1$ cannot take negative values due to the normalization condition \eqref{7.Gnorm}. Therefore, the present model predicts a reduction of power at large scales.
This is a notable difference from the power enhancement obtained in \cite{bib:brizuela-kiefer-2016-desitter,bib:venturi-2013-spettro}, while it agrees with the results in \cite{bib:venturi-2014-spettro-slowroll}.
Actually, it has been debated in \cite{bib:kieferkramer-2013-spettro} that the sign of the spectrum correction depends on the initial conditions implemented for $\Omega^{(1)}$. 
In this sense, we stress that our model dictates $c_1>0$ by the requirement that both polarizations contribute with the same weight in \eqref{7.Gnorm}, therefore it represents a specific solution.

This model therefore predicts a quantum correction to the power spectrum of comoving curvature perturbations due to the presence of gravitons fluctuations on a classical minisuperspace background. Such correction, albeit of very small amplitude (scaled by the presence of the WKB parameter $1/M$), could in principle represent a measurable effect of quantum gravity.

\section{Conclusions}\label{sec:conclusions}

In this work, we addressed the subtle question of constructing unitary quantum gravity corrections to QFT on curved spacetime by a Born-Oppenheimer separation between gravity (the slow system component) and matter (the fast component), in which the former is also treated in a WKB formalism. 

A satisfactory formulation for this problem, up to the zero-order approximation in the Planckian parameter, has been provided in \cite{bib:maniccia-montani-antonini}. The crucial difference between this formulation and previous approaches, like \cite{bib:Vilenkin,bib:kiefer-1991,bib:bertoni-finelli-venturi-1996}, consists in the introduction of an average procedure for the matter wave function over the graviton degrees of freedom, i.e. over the quantum gravity phenomenology. 

However, in order to extend this formulation up to the next order, we needed to introduce a new definition of the internal time variable, simply because the construction of the physical clock \emph{à la} Vilenkin is just the undesired source of nonunitarity of the theory, see \cite{bib:kiefer-1991,bib:digioia-2021}.
Thus, we introduced the Kucha\v{r}-Torre fluid \cite{bib:KucharTorre}, coming from a reparametrized covariant procedure of reference frame fixing, with particular reference to the Gaussian gauge. 
This formulation has already been implemented in \cite{bib:maniccia-montani-2022}, demonstrating its capability to generate a coherent unitary dynamics at the order of quantum gravity corrections too. 
Actually, as a result of its belonging to the fast component of the system, we were able to reconstruct a vacuum Hamilton-Jacobi equation for the classical gravitational field, without any additional nonphysical contributions of the fluid that would violate the so-called energy condition. 
Despite this success, the analysis in \cite{bib:maniccia-montani-2022} still contained, as shown in \cite{bib:maniccia-montani-torcellini}, a particular class of solutions for which the quantum-gravity induced corrections are unable to affect QFT since they take the form of a (time-dependent) factorized phase term. 

In the present study, we recognized the necessity to explicitly express the quantum gravity degrees of freedom, i.e. slow gravitons, as independent variables, on the same level as in \cite{bib:maniccia-montani-antonini}. 
At the zero order of expansion, we were still able to reconstruct the standard QFT on curved space-time by a suitable averaging procedure over the gravitons. 
An important point was that this average is reached by a natural coincidence between the required Born-Oppenheimer gauge fixing and the correct equation for gravity at the first WKB order \cite{bib:barvinsky-kiefer-1998}.
This picture was a solid starting point to perform the further step of calculus, which is the one containing quantum gravity corrections. 
At such order the dependence of the matter sector on the gravitons can no longer be removed, as expected and physically coherent; therefore, the pure QFT state can be provided only as a wave function averaged \emph{a posteriori}. 
We remark that such dynamics is not only a unitary formulation, which does not contain the separable solution giving a net phase outlined in \cite{bib:maniccia-montani-torcellini}. 

We consider the obtained equation to describe the graviton effect on the matter wave function as a fundamental step in shedding light on the emergence of quantum gravity as a purely perturbative phenomenon. 
In order to extract physical information from our model, we implemented the obtained theory to calculate the spectrum of the inflationary perturbations \cite{bib:weinberg-grav-cosm} during a primordial and exact de Sitter phase of the Universe.
The calculus follows a formulation similar to the study in \cite{bib:brizuela-kiefer-2016-desitter} for what concerns the machinery of cosmological perturbations, but the resulting dynamics differs in the time definition and presence of slow graviton contributions. 
The result we obtained for the predicted spectral form is a very small noninvariant correction, see Eq.~\eqref{eq:spettro-fin-marco}, controlled by the ratio between the Hubble constant of the de Sitter phase and the Planckian corresponding scale, i.e., the expansion parameter of the model. 

It is worth stressing that this same result was derived in previous approaches \cite{bib:casadio-2006-spettro,bib:venturi-2013-spettro,bib:brizuela-kiefer-2016-desitter,bib:venturi-2014-spettro-slowroll,bib:venturi-2015-spettro-confronto,bib:brizuela-kiefer-2016-slow-roll,bib:venturi-2016-spettro} (also outside of a pure de Sitter phase) by removing the non-physical terms causing the non-unitary behavior. 
Here, such terms are not present \emph{ab initio} and this fact suggests that we are tracing a viable approach to perturbative quantum gravity, able to provide a similar phenomenology to that one derived in \cite{bib:MVP-2012} but in a consistent and regularized formulation.

\appendix
\section{COMPATIBILITY OF THE GAUSSIAN ANSATZ}\label{appendix}

Here we prove that the Gaussian ansatz \eqref{7.newGauss} and its associated normalization conditions are compatible  with the matter dynamics \eqref{7.mattereq}. Therefore, our task is to determine solutions to the remaining Eqs. \eqref{7.newphi0eq}--\eqref{7.newGeq} of Sec.~\ref{ssec5:gaussian-asatz}.

Let us substitute the solutions \eqref{7.Omega0} and \eqref{7.Nnorm} into \eqref{7.newphi0eq}, which can be solved for $\Upsilon_\textbf{k}^{(0)}$ yielding 
\begin{equation}
    \Upsilon_\textbf{k}^{(0)}(\eta) = - \frac{1}{6}k^3\eta^3\,.
\end{equation}
Thus only Eq.~\eqref{7.newGeq} remains to be solved. We can rewrite it as two equations, one for each polarization state ($v_\textbf{k}^+$ and $v_\textbf{k}^\times$), by decomposing $G_\textbf{k}$ and $\Upsilon_\textbf{k}^{(1)}$ into the following two functions: 
\begin{gather}
    G_\textbf{k}(\eta,v_\textbf{k})=G_{\textbf{k},+}(\eta,v_\textbf{k}^+) + G_{\textbf{k},\times}(\eta,v_\textbf{k}^\times)\,, \label{7.GGlambda}\\
    \Upsilon_\textbf{k}^{(1)}(\eta,v_\textbf{k}) = \Upsilon_{\textbf{k},+}^{(1)}(\eta,v_\textbf{k}^+) + \Upsilon_{\textbf{k},\times}^{(1)}(\eta,v_\textbf{k}^\times)\,. \label{7.phiphilambda}
\end{gather}
This separation does not specify how to distribute the source term $N_\textbf{k} \Omega_\textbf{k}^{(1)}$ in \eqref{7.newGeq} between the two. We consider a reasonable assumption that both polarization states act identically, i.e., they obey the same equations, which is achieved by splitting the source term evenly between the two. We recall that $G_\textbf{k}$ and $\Upsilon_\textbf{k}^{(1)}$ are real by definition \eqref{7.newGauss}, so each (complex) equation of the form \eqref{7.newGeq} for both polarizations can be divided into real and imaginary parts, yielding
\begin{align}
   	\partial_\eta G_{\textbf{k},\lambda} &= - \frac{1}{6}k^3\eta^5 - \frac{1}{2} \partial_\vkl^2 \Upsilon_{\textbf{k},\lambda}^{(1)} - \frac{\vkl}{\eta} \partial_\vkl \Upsilon_{\textbf{k},\lambda}^{(1)} \label{7.dGeq}, \\
	- \partial_\eta \Upsilon_{\textbf{k},\lambda}^{(1)} &= \frac{1}{4} c_1 \eta^2 - \frac{1}{2} \partial_\vkl^2 G_{\textbf{k},\lambda} - \frac{\vkl}{\eta} \partial_\vkl G_{\textbf{k},\lambda}\,, \label{7.dphieq}
\end{align}
where we have used the solutions \eqref{7.Omega0}, \eqref{7.Omega1}, \eqref{7.Nnorm} and computed \eqref{7.xi} in the super-Hubble limit, namely 
\begin{equation}
    \xi_k(\eta) \sim - \frac{1}{\eta}.
\end{equation}
Indeed, while $\Re(\xi_k(\eta))$ is divergent for $\eta \rightarrow 0^-$, the imaginary part is infinitesimal, so that the dominant contribution is the real one.

We look for solutions to Eqs.~\eqref{7.dphieq}-\eqref{7.dGeq} by considering each component in \eqref{7.GGlambda}-\eqref{7.phiphilambda} of the form
\begin{align}
    \Upsilon(\eta,\vkl) &= \bar{\Upsilon}(\eta) \exp \left[ -\frac{1}{2}\sigma_\Upsilon (\vkl)^2 \right], \label{7.phiGauss} \\
    G(\eta,\vkl) &= \bar{G}(\eta) \exp \left[ - \frac{1}{2} \sigma_G(\eta) (\vkl)^2 \right], \label{7.GGauss}
\end{align}
where all the subscripts and superscripts are implied. Note that while $\sigma_\Upsilon$ is a free parameter, $\sigma_G$ is taken as a function of conformal time. 
From the decomposition \eqref{7.GGlambda} of $G_\textbf{k}$ into the two polarizations, it is useful to observe that
\begin{equation}
   1 + \frac{\Hubble_0^2}{M}G_\textbf{k} = \left( 1 + \frac{\Hubble_0^2}{M} G_{\textbf{k},+} \right) \left( 1 + \frac{\Hubble_0^2}{M} G_{\textbf{k},\times} \right)
\end{equation}
is valid up to $\order{M^{-1}}$ and can be used in \eqref{7.newGauss}. Clearly, $G_{\textbf{k},+}$ and $G_{\textbf{k},\times}$ take the same functional form in the hypothesis of even splitting between the two polarizations, so the condition \eqref{7.Gnorm} can be recast using \eqref{7.Omega0}-\eqref{7.Omega1} as
\begin{equation}
    \int d\vkl \ G_{\textbf{k},\lambda} = \sqrt{\frac{c_1 }{4k^3}}\,.
\end{equation}
The factor $\bar{G}(\eta)$ in \eqref{7.GGauss} is immediately determined as
\begin{equation}
    \bar{G}(\eta) = \sqrt{\frac{c_1\,\sigma_G(\eta)}{8\pi k^3}}\,.
	\label{7.barG}
\end{equation}
Now plugging \eqref{7.phiGauss}--\eqref{7.GGauss} into \eqref{7.dGeq}--\eqref{7.dphieq} we obtain ordinary differential equations that, neglecting all terms quadratic in the $\vkl$, give
\begin{gather}
   \frac{d}{d\eta} \sqrt{\sigma_G} = \sqrt{8\pi k^3} \left(\frac{1}{2} \sigma_\Upsilon \bar{\Upsilon}- \frac{\sqrt{c_1}}{6} k^3 \eta^5 \right), \label{7.wGeq} \\
	- \frac{d}{d\eta} \bar{\Upsilon} = \sqrt{\frac{c_1\, \sigma_G^{3}}{8\pi k^3}}  + \frac{1}{4} c_1\eta^2. \label{7.barphieq}
\end{gather}

So far, we have reduced the problem to a a system of two coupled ordinary differential equations. 
We can determine an explicit solution in our limit of interest $\eta \rightarrow 0^-$ by performing a series expansion in $\eta$:
\begin{gather}
    \sigma_G(\eta) = A_0 + A_1 \eta + \order{\eta^2}, \\
    \bar{\Upsilon}(\eta) = B_0 + B_1 \eta + \order{\eta^2},
\end{gather}
with $A_0$, $A_1$, $B_0$, $B_1$ constants depending on the initial conditions.
Substituting into Eqs.~\eqref{7.wGeq}-\eqref{7.barphieq} and neglecting all terms of $\order{\eta^2}$ or higher we find
\begin{gather}
    A_1 = \sqrt{\frac{8\pi k^3 A_0}{c_1}}\,\sigma_\Upsilon\, B_0\,, \\
	B_1 = -\sqrt{\frac{c_1}{8\pi}} \left( \frac{A_0}{k} \right)^{3/2}\,,
\end{gather}
leading to the (approximate) functions
\begin{gather}
   \sigma_G(\eta) = A_0 +\sqrt{\frac{8\pi k^3 A_0}{c_1}}\,\sigma_\Upsilon\, B_0\,\eta\,, \label{7.wG} \\
	\bar{\Upsilon}(\eta) = B_0 - \sqrt{\frac{c_1}{8\pi}} \left( \frac{A_0}{k^3} \right)^{3/2} \eta\,\,, \label{7.barphi}
\end{gather}
where $A_0$, $B_0$, and $\sigma_{\Upsilon}$ must be positive to obtain well-defined solutions. It is clear from \eqref{7.wGeq}--\eqref{7.barphieq} that $A_0 \equiv \sigma_G(0)$ and $B_0 \equiv \bar{\Upsilon}(0)$, i.e., they represent the initial conditions of the system.

Now inserting \eqref{7.wG}--\eqref{7.barphi} and \eqref{7.barG} into \eqref{7.phiGauss}--\eqref{7.GGauss}, we conclude that there exists a solution to Eq.~\eqref{7.mattereq} in the form of a normalized Gaussian \eqref{7.newGauss} and this validates the choice implemented in Sec.~\ref{ssec5:gaussian-asatz}. 
Regardless, the functions \eqref{7.wG}--\eqref{7.barphi} here obtained are redundant since they do not enter the computation of the modified power spectrum in the super-Hubble limit.

\bibliography{Articolo}

\end{document}